\documentclass[onecolumn,12pt]{article}
\usepackage{jheppub}
\usepackage{ifpdf}
\usepackage[bb=boondox]{mathalfa}
\usepackage{graphicx,subcaption}
\graphicspath{ {figures/} }
\usepackage{amsfonts}
\usepackage{amsmath,braket}
\usepackage{amssymb}
\usepackage{epsfig}
\usepackage{tensor}
\usepackage{pdfpages}
\usepackage{bbm}
\usepackage{graphicx,epstopdf}
\usepackage[numbers]{natbib} 
\usepackage[makeroom]{cancel}
\usepackage{hyperref}
\usepackage{array}
\usepackage[export]{adjustbox}

\usepackage[normalem]{ulem}
\usepackage{romannum}

\usepackage{ulem}

\usepackage{color}
\usepackage{here}

\numberwithin{equation}{section}									% equation numbering by section

\usepackage{comment}

\newcommand{\de}{\partial}
\newcommand{\be}{\begin{equation}}
	\newcommand{\ba}{\begin{eqnarray}}
		\newcommand{\ea}{\end{eqnarray}}
	\newcommand{\ee}{\end{equation}}

\newcommand{\f}{\frac}
\newcommand{\s}{\sqrt}
\newcommand{\vp}{\varphi}

\newcommand{\ap}{\alpha}
\newcommand{\pr}{^{\prime}}
\newcommand{\prpr}{^{\prime\prime}}
\newcommand{\prsq}{^{\prime2}}

\newcommand{\ddd}{\cdot\cdot\cdot}
\newcommand{\no}{\nonumber \\}

\newcommand{\bea}{\begin{eqnarray}}
	\newcommand{\eea}{\end{eqnarray}}
\newcommand{\bes}{\begin{equation*}}
	\newcommand{\beas}{\begin{eqnarray*}}
		\newcommand{\eeas}{\end{eqnarray*}}
	\newcommand{\bas}{\begin{array*}}
		\newcommand{\eas}{\end{array*}}
	\newcommand{\ees}{\end{equation*}}

\newcommand{\ep}{\epsilon}

\let\a=\alpha \let\b=\beta   \let\e=\epsilon  \let\g=\gamma  \let\k=\kappa \let\l=\lambda \let\m=\mu 
  \let\r=\rho %\let\s=\sigma
\let\t=\tau   \let\vp=\varphi   
 \let\D=\Delta \let\G=\Gamma \let\L=\Lambda   \let\S=\Sigma \let\Th=\Theta

\newcommand{\abs}[1]{\vert #1 \vert}

\newcommand{\eps}{\epsilon}

\newcommand{\Poincare}{Poincar\'{e} }

\subheader{\today}
\title{\boldmath Brane Cosmology from AdS/BCFT}

\author[a]{Kosei Fujiki,}
\author[a]{Hiroki Kanda,}
\author[a]{Michitaka Kohara,}
\author[a,b]{Tadashi Takayanagi}

\affiliation[a]{Center for Gravitational Physics and Quantum Information, Yukawa Institute for Theoretical Physics, Kyoto University,\\/
	Kitashirakawa Oiwakecho, 
    Sakyo-ku, Kyoto 606-8502, Japan}
\affiliation[b]{Inamori Research Institute for Science,\\
	620 Suiginya-cho, Shimogyo-ku,Kyoto 600-8411 Japan}

\emailAdd{kosei.fujiki@yukawa.kyoto-u.ac.jp}
\emailAdd{hiroki.kanda@yukawa.kyoto-u.ac.jp}
\emailAdd{michitaka.kohara@yukawa.kyoto-u.ac.jp}
\emailAdd{takayana@yukawa.kyoto-u.ac.jp}

\abstract{In this paper, we study the time-dependent dynamics of an end-of-the-world (EOW) brane in AdS with a scalar field localized on the brane. We mainly studied several aspects of holography and cosmology.
Standard requirements in the AdS$_{d+1}$/CFT$_d$ lead to a constraint on the conformal dimension in the dS$_d$/CFT$_{d-1}$. We also prove a time-like analog of g-theorem using the null energy condition in the context of AdS$_3$/BCFT$_2$.
In the cosmological interpretation, we rewrite the equation of motion of the brane as a Friedman-like equation, which enables us to consider its dynamics in analogy with the ordinal cosmology. And then we classify all possible solutions of the brane when the potential takes a constant value. We find that our brane cosmology model can describe a process of creating a universe via a big-bang. Additionally, we show that when the brane is close to a hyperplane, its effective action is given by a Liouville gravity with a scalar field matter. Finally, we 
also obtain brane solutions with boost symmetry, which are obtained by analytical continuation of Euclidean branes with a torus topology.
}

\begin{document} 
	
	%%%%%%%%%%%%%%%%%%%%%%%%%%%%%%%%%%%%%%%%%%%%%%%%%%%%
	\begin{flushright}
		YITP-24-180
        \\
	\end{flushright}
	%%%%%%%%%%%%%%%%%%%%%%%%%%%%%%%%%%%%%%%%%%%%%%%%%%%%
	\maketitle
	\flushbottom

%%%%%%%%%%%%%%%%%%%%%%%%%%%%%%%%%%%%%%%%%%%%%%%%%%%%%%%%
%%%%%%%%%%%%%%%%%%%%%%%%%%%%%%%%%%%%%%%%%%%%%%%%%%%%%%%%
\section{Introduction}
\label{sec:intro}
%%%%%%%%%%%%%%%%%%%%%%%%%%%%%%%%%%%%%%%%%%%%%%%%%%%%%%%%
%%%%%%%%%%%%%%%%%%%%%%%%%%%%%%%%%%%%%%%%%%%%%%%%%%%%%%%%

The end-of-the-world (EOW) brane has recently played crucial roles in the AdS/CFT correspondence \cite{Maldacena:1997re,Gubser:1998bc,Witten:1998qj}. First of all, it provides a gravity dual of a conformal field theory (CFT) on a manifold with boundary i.e. the boundary conformal field theory (BCFT), so called the AdS/BCFT correspondence \cite{Takayanagi:2011zk,Fujita:2011fp,Karch:2000gx}. Moreover, EOW branes provide many new insights on the quantum gravity aspects of holography. For example, it provides a class of excellent models \cite{Almheiri:2019hni,Almheiri:2019psf} to explore black hole information problem in the light of island formula \cite{Penington:2019npb,Almheiri:2019psf}. Indeed, the computation rule of holographic entanglement entropy \cite{Ryu:2006bv,Ryu:2006ef,Hubeny:2007xt} is modified in the presence of EOW branes \cite{Takayanagi:2011zk,Fujita:2011fp} and this can be understood a special form of the island formula \cite{Suzuki:2022xwv,Izumi:2022opi}. EOW branes have also provided us with useful approached to constructing models for quantum cosmology \cite{Cooper:2018cmb,Antonini:2019qkt,VanRaamsdonk:2021qgv,Omiya:2021olc,Antonini:2022blk,Waddell:2022fbn} by combining the AdS/BCFT with the brane-world holography \cite{Randall:1999ee,Randall:1999vf,Gubser:1999vj,Shiromizu:1999wj,Garriga:1999yh,Karch:2000ct,Shiromizu:2001jm,Koyama:2001rf,Shiromizu:2001ve}. EOW branes have also been an important tool to build holographic models of condensed matter systems 
\cite{Fujita:2012fp,Gutperle:2012hy,Hartman:2013qma,Estes:2014hka,Erdmenger:2014xya,Erdmenger:2015spo,Seminara:2018pmr,Shimaji:2018czt,Kanda:2023zse,Kanda:2023jyi}.

\begin{figure}[htbp]
		\centering
		\includegraphics[width=13cm]{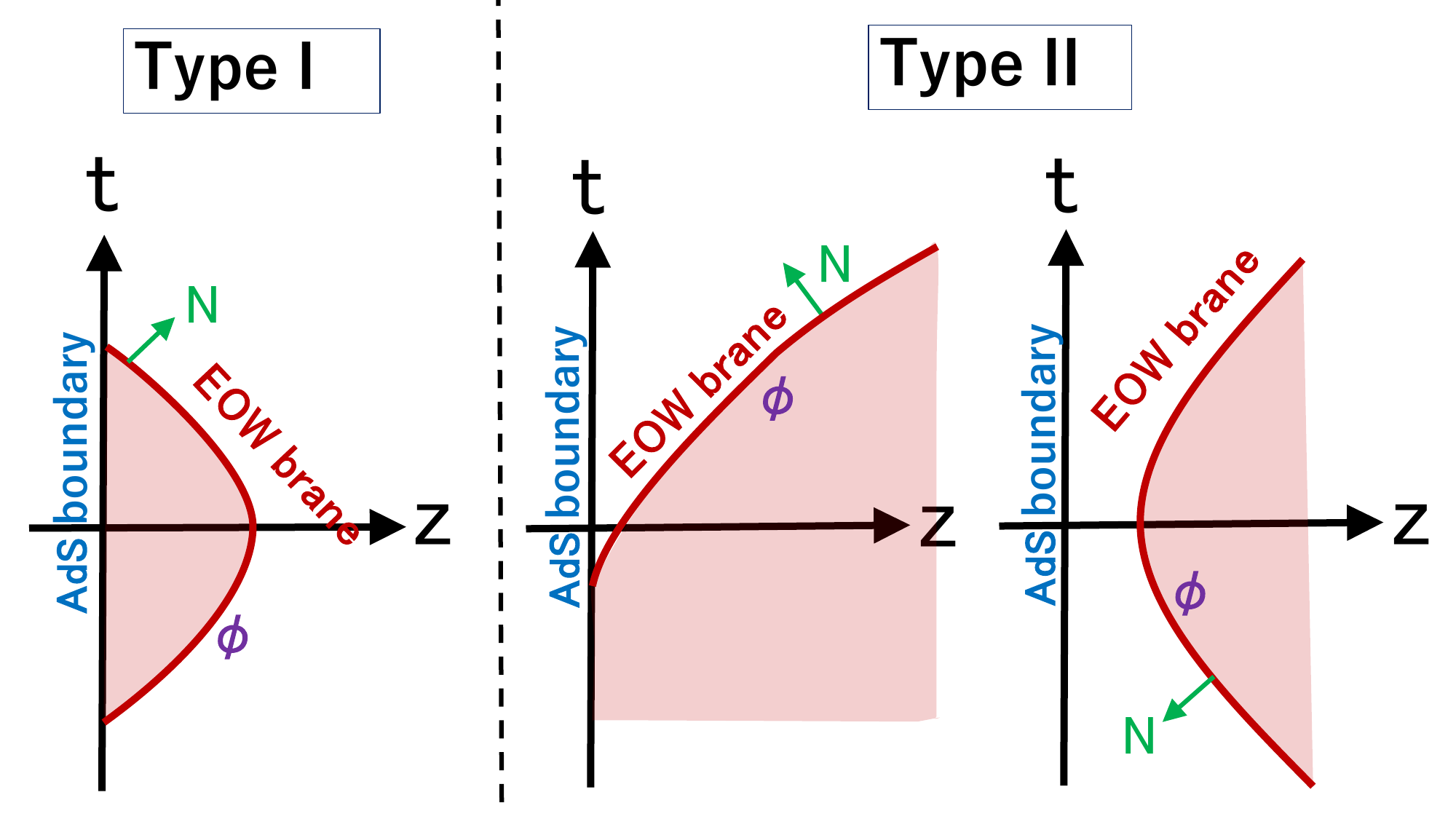}
		\caption{Two setups: type I (left) and type II (middle and right) of EOW branes in AdS/BCFT. The radial and time coordinate of AdS are denoted by $z$ and $t$, respectively, such that the AdS boundary is situated at $z=0$. The normal vector $N$, for which we calculate the extrinsic curvature, is also indicated.} 
		\label{fig:setup}
\end{figure}

In most of previous studies of EOW branes and AdS/BCFT, purely gravity models were assumed in order to analytically solve the dynamics of the brane coupled to the AdS gravity. Thus, the effects of matter fields have not been well understood especially in time-dependent backgrounds, in spite that matter fields are important ingredient to construct various cosmological models. Motivated by this, we would like to analyze a class of AdS/BCFT models with a scalar field localized on an EOW brane, which we can often solve analytically. This class was first studied in \cite{Kanda:2023zse,Kanda:2023jyi} and various static brane solutions which are dual to BCFTs with time-like boundaries were found. This includes gravity duals of boundary renormalization group (RG) flows and in particular an interesting phase transition which is very similar to the entanglement phase transition was realized.

In this paper, we will employ the same model mainly focusing on a time-like EOW brane with a localized scalar field $\phi$ in AdS$_3$ and will analyze various time-dependent solutions of EOW branes which either do not end on the AdS boundary or do end on a space-like region. We concentrate on the brane with the translational invariance in space-like directions, which excludes well-studied AdS branes. This model has an advantage that we can solve its back reaction easily because any solution in the three dimensional pure gravity coincides with the AdS$_3$ locally. This simple model turns out to have a rich variety of cosmological solutions. Even though the bulk spacetime is an anti de Sitter space (AdS), the brane spacetime can be a de Sitter (dS) space, which has a positive cosmological constant. A holography in de Sitter space, so called dS/CFT \cite{Strominger:2001pn,Witten:2001kn,Maldacena:2002vr}, has still been not well understood mainly because the dual CFTs are expected to be highly exotic \cite{Anninos:2011ui,Hikida:2021ese}. Therefore, our model provides another route to study the dS/CFT by embedding it in a higher dimensional AdS/CFT. Even though we will concentrate on the lowest dimensional model i.e. AdS$_3/$BCFT$_2$, we can extend our analysis to higher dimensional AdS/BCFT, though in general it is not straightforward to solve the back reaction problem. We would also like to mention that dS branes have another interesting application to holography in the context of gravity duals of CFTs on non-oriented manifolds \cite{Wei:2024zez,Wei:2024kkp}.

In general, the spacetime with a EOW brane can be classified into two types: type I and type II. For a type I brane, the bulk spacetimes is the region surrounded by the brane and the AdS boundary as depicted in the left panel of Fig.\ref{fig:setup}. On the other hand, for a type II brane, the bulk region is the interior side with respect to the brane. It can be either a region which is surrounded by the brane and the AdS boundary (i.e. middle panel of Fig.\ref{fig:setup}) or a completely interior region which does not reach the AdS boundary as in the right panel of the figure. 

\begin{figure}[!htbp]
        \centering
        \includegraphics[width=7cm]{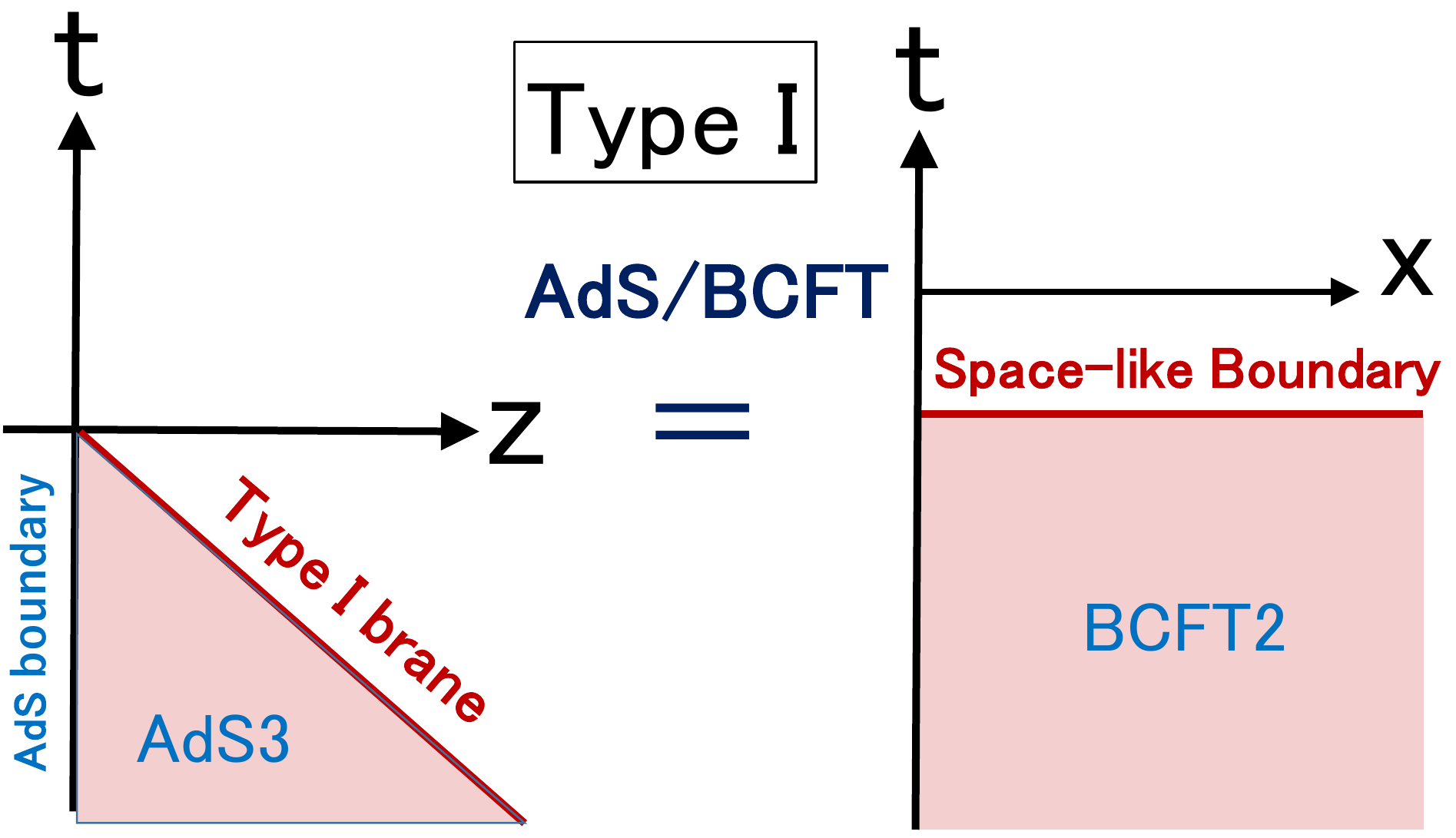}
        \hspace{1cm}
        \includegraphics[width=7cm]{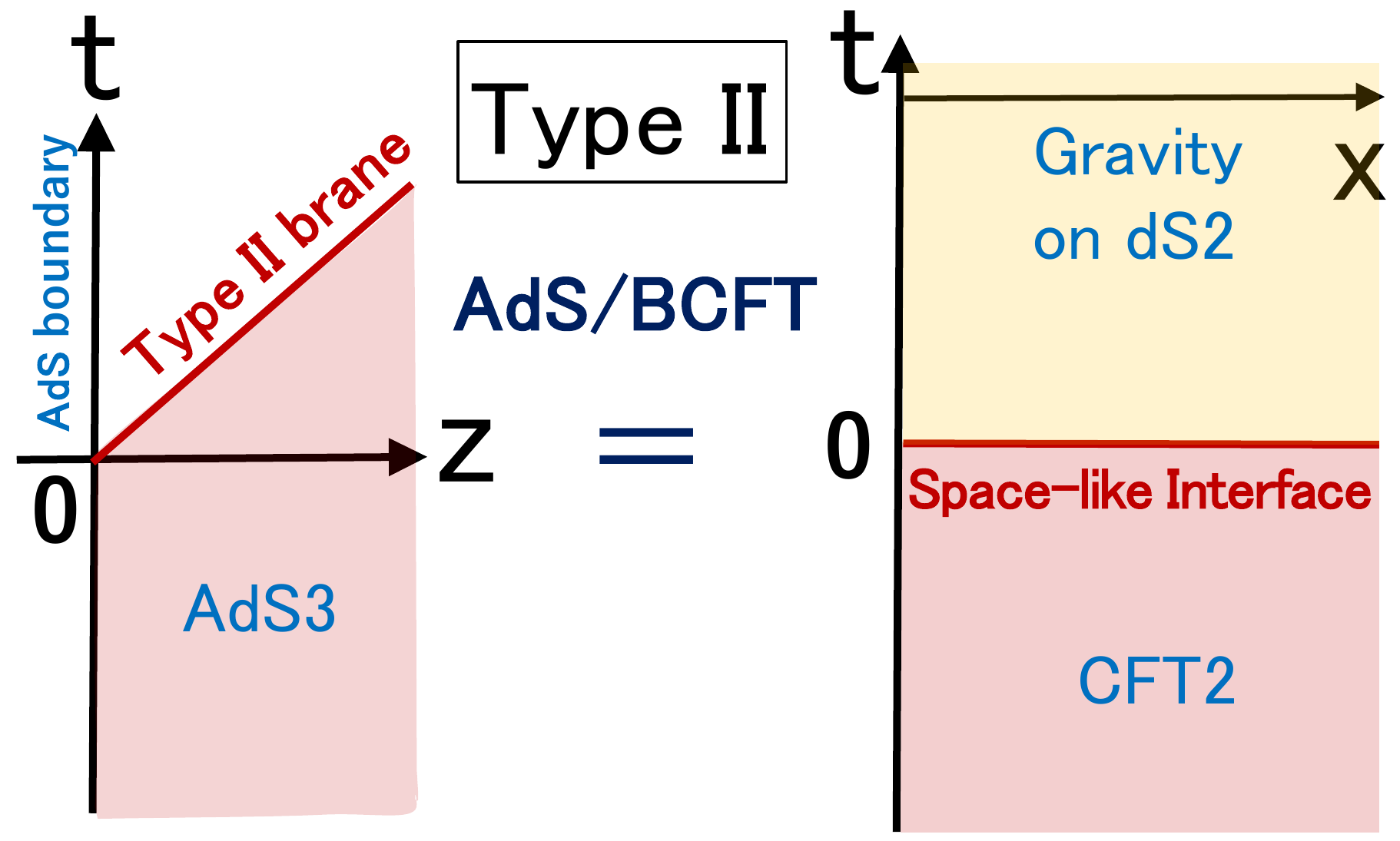}
        \caption{DS$_2$ brane and its CFT dual in the type I (left) and type II (right) case. They are solutions when the scalar field takes a constant value. If we place BCFT on upper half plane, our result is applicable.} 
        \label{fig:hyperplane}
\end{figure}

Before we turn on the localized scalar, the simplest solution is given by the setup where the EOW brane is a time-like hyperplane, as shown in Fig.\ref{fig:hyperplane}.
This hyperplane is identical to two dimensional de Sitter space (dS$_2$) in the bulk AdS$_3$. In the type I case, the bulk gravity is expected to be dual to a two dimensional BCFT with the space-like boundary at $t=0$ \cite{Akal:2020wfl}. We can view this as a CFT with the final state projection at $t=0$ \cite{Akal:2021dqt,Numasawa:2016emc}. If we turn on the scalar field on the EOW brane, then we can describe the boundary RG flow in the presence of the space-like boundary. We have been lacking knowledge of quantum field theory with a space-like boundary, because it has unusual properties such as the causality violation. Our AdS/BCFT model provides a powerful tool to explore such a setup. Indeed, we will manage to prove a monotonicity property analogous to $g$-theorem from our holographic description.

In the type II case, which is more interesting from the cosmological interpretation, the AdS$_3$ gravity with the EOW brane which is given by the two dimensional hyperplane is dual to a two dimensional system which consists of a CFT in the lower half plane and a gravity on dS$_2$, which are coupled with each other at the time slice $t=0$ \cite{Chen:2020tes,Akal:2020wfl}. Since they are dual to the bulk gravity, the gravity on dS$_2$ should include quantum gravity effects. By turning on the scalar field on the brane, we can study the dynamics of quantum gravity on dS$_2$, from which we will get implications on the dS/CFT. In the presence of scalar field, we can construct type II branes which do not reach the AdS boundary. By applying the brane world holography, we expect that the three dimensional bulk gravity inside the brane is dual to the two dimensional quantum gravity on the EOW brane. 
This provides us with a simple toy model of quantum cosmology. Indeed we will be able to find various interesting cosmological models where the Universe is created at a time and expands like a big-bang process. Depending on the values of potential of the scalar field on the brane, it can either continue to expand and start to shrink. 

This paper is organized as follows. In section \ref{sec:dSBranes}, we describe our basic setup of an EOW brane with a localized scalar and present several interesting solutions. This includes perturbed de Sitter branes and we will discuss implications of our results on the dS/CFT correspondence. In section \ref{sec:g-th}, we consider boundary RG flows in our holographic models and derive an analogue of holographic $g$-theorem. 
In section \ref{sec:Liouville}, we argue that the dynamics of EOW branes is well approximated by a Liouville gravity coupled to a scalar field when the brane fluctuations are small.
In section \ref{sec:cosmology}, we analyze the cosmological aspects of EOW brane dynamics. We also classify time-dependent solutions when the potential is a constant. In section \ref{sec:boost}, we study the case where EOW branes have the boost symmetry. In section \ref{sec:conclusions}, we summarized our conclusions and discuss future problems. In appendix \ref{ap:Vzero}, we present analytical solutions when the potential is vanishing. In appendix \ref{ap:Vnonzero}, analytical solutions for a non-zero constant potential are derived. In appendix \ref{sec:EuclidS}, the Euclidean brane solutions are computed and classified. In appendix \ref{ap:hyp}, branes in the hyperbolic slice of AdS are analyzed.

%%%%%%%%%%%%%%%%%%%%%%%%%%%%%%%%%%%%%%%%%%%%%%%%%%%%%%%%
%%%%%%%%%%%%%%%%%%%%%%%%%%%%%%%%%%%%%%%%%%%%%%%%%%%%%%%%
\section{dS Branes in AdS$_3$}
\label{sec:dSBranes}
%%%%%%%%%%%%%%%%%%%%%%%%%%%%%%%%%%%%%%%%%%%%%%%%%%%%%%%%
%%%%%%%%%%%%%%%%%%%%%%%%%%%%%%%%%%%%%%%%%%%%%%%%%%%%%%%%
 As in \cite{Kanda:2023zse,Kanda:2023jyi}, we assume the AdS/BCFT setup with a localized scalar $\phi$ on the EOW brane described by the action:
\begin{align}\label{eq:BraneScalarLagrangian}
I=&\frac{1}{16\pi G_N}\int d^3x \s{-g}(R-2\Lambda)+\frac{1}{8\pi G_N}\int_\Sigma d^2x\s{-h}K  \\
&+\frac{1}{8\pi G_N}\int_Q d^2x\s{-h}\left(K-h^{ab}\de_a\phi\de_b\phi-V(\phi)\right),\nonumber
\end{align}
where $\Lambda=-1$ in the current convention, $\Sigma$ is the asymptotic boundary of the AdS$_3$ spacetime, and $Q$ is a time-like two dimensional hypersurface  and is called the end-of-the-world brane (EOW) brane. The variation of the action reads 
\ba
\delta I=\frac{1}{16\pi G_N}\int_Q d^3x \s{-h}\delta h^{ab}\left[
K_{ab}-h_{ab}K-2\de_a\phi\de_b\phi+h_{ab}\left(h^{cd}\de_c\phi\de_d\phi+V(\phi)\right)\right],\no
\ea
where $h_{ab}$ and $K_{ab}$ are the induced metric and the extrinsic curvature on the surfaces $\Sigma$ and $Q$, where we chose the normal vector in the outward direction. Its trace is expressed as $K=h^{ab}K_{ab}$. 

By setting this variation to zero, we find that the Neumann boundary condition on the EOW brane is 
\begin{align}
&K_{ab}-h_{ab}K=T_{Q\,ab},\label{KEOM} \\
 & \mbox{where},\ T_{Qab}\equiv 2\de_a\phi\de_b\phi-h_{ab}\left(h^{cd}\de_c\phi\de_d\phi+V(\phi)\right).  
\end{align}

The equation of motion of the localized scalar reads
\ba
2\de_a(\s{-h}h^{ab}\de_b\phi)-\s{-h}V'(\phi)=0.  \label{SEOM}
\ea
It is useful to note that the above EOM for the scalar can be found from (\ref{KEOM}) by noting $\nabla^a(K_{ab}-h_{ab}K)=0$ in the pure AdS$_3$.
Therefore, in the following, we consider the solutions to (\ref{KEOM}) in AdS$_3$
(\ref{PAdS}).

It is also very important and convenient to note that all the solutions to the vacuum Einstein equation with the negative cosmological constant $R_{\mu\nu}+2g_{\mu\nu}=0$ are given by pure AdS$_3$ (\ref{PAdS}) locally. Therefore, we do not need to worry about the backreaction due to the EOW brane, which is present in higher dimensions.

For simplicity we assume that the scalar field depends only on some time-like parameter\footnote{Depending on the profile of the brane, it can be space-like parameter.} 
\begin{align}
    \phi=\phi(t)
\end{align}
In this case the null-energy-condition (NEC) on EOW is automatically satisfied as long as the scalar field takes real values, which is easily confirmed as follows.
Without loss of generality, we can set the induced metric in a form
\begin{align}
    ds^2=h_{tt}dt^2+h_{xx}dx^2.  
\end{align}
Here, $x$ denotes any other space-like coordinate: $x$ for Poincar\'{e}, $\phi$ for global coordinate.
The arbitral null vector is $k=C(\sqrt{h_{xx}}\partial_t+\sqrt{-h_{t t}}\partial_x)$, and we set $C=1$.
The components of the energy-momentum tensor is
\begin{align}
    T_{Qtt}&=\dot{\phi}^2-h_{tt}V(\phi) \\
    T_{Qxx}&=-h_{xx}(h^{tt}\dot{\phi}^2+V(\phi)) \label{EMtensor}
\end{align}
Thus NEC is
\begin{align}
    T_{Q ab}k^ak^b
    &=(\dot{\phi}^2-h_{tt}V(\phi))h_{xx}+(-h_{xx})(h^{tt}\dot{\phi}^2+V(\phi))(-h_{tt})\\
    &=2h_{xx}\dot{\phi}^2 \geqq 0.
\end{align}

In this paper, the extrinsic curvature is always diagonal $K_{tx}=0$, in such cases we can solve the Neumann boundary condition \eqref{KEOM} for $\dot{\phi}^2$ and $V$. The result is 
\begin{align}
    \dot{\phi}^2&=\f{1}{2}(K_{tt}-h_{tt}h^{xx}K_{xx}) \\
    V&= \f{1}{2}K \label{phid2andV}
\end{align}
Note that the second equation is just the trace of eq \eqref{KEOM}, so it holds in any situation.

\subsection{General argument in Poincar\'{e} AdS$_3$}
\label{subsec:generalArgInAdS3}
In this section, we choose the Poincar\'{e} coordinate of AdS$_3$ and analyze the EOW brane:
\begin{align}
    ds^2=\frac{dz^2-dt^2+dx^2}{z^2}.  \label{PAdS}
\end{align}
We insert an EOW brane along the trajectory $z=Z(t)$, which is also extended in the $x$ direction. There are two different setups of AdS/BCFT, depicted in Fig.\ref{fig:setup}:
\ba
&& \mbox{Type I:}\ \ \ 0\leq z\leq Z(t),\no
&& \mbox{Type II:}\ \ \ z\geq Z(t).
\ea
In the type I case, the physical region in three-dimensional gravity extends from the AdS boundary to the EOW brane. As we shall see, the type I EOW brane always intersects with the AdS boundary.
However, in the type II case, the physical region is inside the EOW brane. The type II EOW brane either stays entirely within the interior of AdS$_3$ or intersects the AdS boundary.

The induced metric on the EOW brane reads
\begin{align}
    ds^2_{Q}=\frac{-(1-\dot{Z}^2)dt^2+dx^2}{Z^2},
\end{align}
where $\dot{Z}=\f{dZ}{dt}$.
In order to make the brane time-like, we impose $1-\dot{Z}^2>0$. 
The normal vector is 
\begin{align}
    N=\epsilon\frac{dz-\dot{Z}dt}{Z\sqrt{1-\dot{Z}^2}} \label{P_normal}
\end{align}
where $\epsilon=\pm 1$. When $\eps=1$ (or $\eps=-1$), $N$ has a positive (or negative) direction on the $z$ axis, which is a type I (or type II) set up.  This expression of normal vector leads to 
\begin{align}
  K_{ab}dx^adx^b=\eps\frac{1-\dot{Z}^2-Z\ddot{Z}}{Z^2\s{1-\dot{Z}^2}}dt^2-\eps\frac{1}{Z^2\s{1-\dot{Z}^2}}dx^2.
\end{align}
The boundary condition (\ref{phid2andV}) is explicitly written as follows:
\begin{align}
    \dot{\phi}^2&=-\frac{\eps\ddot{Z}}{2Z\s{1-\dot{Z}^2}} \label{phid2_Z}\\
    V&=\frac{\eps Z\ddot{Z}}{2(1-\dot{Z}^2)^{3/2}}-\frac{\eps}{\sqrt{1-\dot{Z}^2}}. \label{V_Z}
\end{align}
% Type Iの場合はV<-1になる
We immediately see that the null energy conditions are $\ddot{Z} \leq 0$ in type I and $\ddot{Z} \geq 0$ in type II. This shows that the type I brane always intersects with the AdS boundary in the past and future, which is related to the fact that the potential $V$ always satisfies $V<-1$ for the type I brane. On the other hand, the potential $V$ of the type II brane has neither upper nor lower bound.

There are some setups which are obviously prohibited by the null energy condition. Consider two EOW branes such that the bulk sub-region enclosed by them connects the two regions $[t_1,t_2]$ and $[t_3,t_4]$ on the asymptotic boundary. We can interpret this as a kind of wormhole, which extends in the time-like direction as opposed to usual space-like ones, from the perspective of BCFT. However, we find that it is not allowed because the inner side of the EOW brane does not satisfy the type II null energy condition. Thus the time-like wormhole is prohibited by the null energy condition. 

% \begin{figure}[htbp]
%         \centering
%         \includegraphics[width=4cm]{wormhole.pdf}
%         \caption{Time-like Wormhole like EOW brane} 
%         \label{fig:wormhole}
% \end{figure}

From (\ref{phid2_Z}) and (\ref{V_Z}), we obtain the kinematic-energy-conservation-like relation;
\begin{align}
    \f{Z^2}{\sqrt{1-\dot{Z}^2}}\dot{\phi}^2+\sqrt{1-\dot{Z}^2}V(\phi)=-\eps.
\end{align}

As an trivial example, we can set a hyperplane 
\be
Z(t)=-\e\lambda t,  \label{dSbranea}
\ee
with time-like condition $0<\lambda<1$. This setup is shown in Fig.\ref{fig:hyperplane}. This solves the boundary condition when $\phi=\mathrm{const}.$ and $V=-\eps/\sqrt{1-\lambda^2}$. Induced metric is 
\begin{align}
    ds_Q^2=\frac{-(1-\lambda^2)dt^2+dx^2}{\lambda^2 t^2}.  
\end{align}
Thus when $Z(t)$ approaches to $-\e \l t + O(t^2)$, the geometry of the brane approaches to dS$_2$ with radius $R_{dS}=\sqrt{1-\lambda^2}/\lambda$ near the asymptotic boundary.

Now let us consider the CFT dual of the type I and type II hyperplane dS$_2$ solution (\ref{dSbranea}). The AdS/BCFT \cite{Takayanagi:2011zk,Fujita:2011fp} argues that the gravity dual of a boundary conformal field theory (BCFT) on a $d$ dimensional manifold $S$ is given by the gravity on the $d+1$ dimensional region $\Sigma$ which boundaries consist of $S$ and an EOW brane $Q$, such that $\de\Sigma=S\cup Q$. The manifold $S$ is assumed to have a boundary $\de S$ and this clearly coincides with $\de Q$. In the Lorentzian AdS, this was originally argued when the boundary $\de S$ is time-like. 

In our setup, on the other hand, $\de Q=\de S$ is space-like i.e. $t=0$ and thus we need to consider a generalization of AdS/BCFT. This was first considered in  \cite{Akal:2020wfl}. As depicted in Fig.\ref{fig:hyperplane}, we expect that the type I brane geometry is dual to a two dimensional BCFT i.e. a two dimensional CFT on the lower half plane. On the other hand, the type II one is dual to a two dimensional system which consists of two dimensional CFT on the lower half plane $t<0$ and the gravity on the dS$_2$ in the upper half plane $t>0$, which are coupled with each other at the space-like interface $t=0$ \cite{Chen:2020tes,Akal:2020wfl}.

\subsection{Perturbations of dS branes and dS$_2/$CFT$_1$}\label{pertds}

Now let us consider a perturbation of the dS$_2$ brane solution (\ref{dSbranea}) by turning on the scalar field $\phi$ on the brane. We assume that the brane asymptotically behaves like:
\begin{align}
    Z(t)\simeq -\alpha t-\eps\beta t^p +o(t^p)  \label{perttoh}
\end{align}
in the $t\to 0$ limit. Since it is time-like and satisfies the null energy condition we require 
\begin{align}
    0<\alpha<1,\ \,\beta>0,\,\ \,p>1
\end{align}
Then \eqref{phid2_Z} and \eqref{V_Z} approach to the following expression 
\begin{align}
    \dot{\phi}^2&\simeq \frac{p(p-1)\beta}{2\ap\s{1-\ap^2}}t^{p-3}, \\
    V(\phi)&\simeq -\frac{\eps}{\s{1-\ap^2}}-\frac{p(p-3)\ap\beta}{2(1-\ap^2)^{3/2}}t^{p-1}.
\end{align}
By solving the first equation we find
\begin{align}
    \phi\simeq \phi_*\pm \s{\frac{2p\beta}{(p-1)\ap\s{1-\ap^2}}}t^{\frac{p-1}{2}}.
\end{align}
Thus the potential can be rewritten as 
\begin{align}
    V(\phi)\simeq -\frac{\eps}{\s{1-\ap^2}}+\frac{\ap^2(p-1)(3-p)}{4(1-\ap^2)}(\phi-\phi_*)^2+\ddd  \label{potoneg}
\end{align}
First of all, this shows that we need $V\leq -1$ in type I
and $V\geq 1$ in type II at the AdS boundary.
% II 特にエンドする点でっていう話

The mass of the scalar field is found from (\ref{potoneg}) to be $M^2=\frac{\ap^2(p-1)(3-p)}{4(1-\ap^2)}$. Since the radius of dS$_2$ before the perturbation reads $R_{dS}^2=\frac{1-\ap^2}{\ap^2}$, we obtain
\begin{align}
    M^2 R_{dS}^2=\frac{(p-1)(3-p)}{4}=\frac{-(p-2)^2+1}{4}\leq \frac{1}{4}.
    \label{ubfgr}
\end{align}

Now let us focus on the type II brane setup, which looks like the right panel of Fig.\ref{fig:hyperplane}. Since the dS-brane is gravitating, we expect that two dimensional gravity is localized on the brane as in the brane-world model
\cite{Randall:1999ee,Randall:1999vf,Karch:2000ct,Gubser:1999vj}. In summary, gravity with the type II brane is dual to a two dimensional CFT on the Lorentzian lower half plane $t<0$ coupled at $t=0$ to a gravitational theory on dS$_2$ extends in the upper half $t>0$. We expect that the latter i.e. gravity on dS$_2$ is dual to a one dimensional conformal quantum mechanics localized on $t=0$ assuming the dS/CFT duality \cite{Strominger:2001pn,Witten:2001kn,Maldacena:2002vr}. Notice that since the quantum mechanical degrees of freedom at $t=0$ is coupled to the two dimensional CFT on $t<0$, the total theory might properly be regarded as a boundary conformal field theory with a space-like boundary. Below we would like to examine the implication of our EOW brane analysis in terms of dS/CFT. As we will discuss later in section \ref{sec:higherds}, we can extend our argument here to 
higher dimensions in a straightforward way.

The scalar field $\phi$ localized on the dS brane can be regarded as the bulk scalar on dS$_2/$CFT$_1$. Thus the scalar field is dual to a scalar operator in the CFT$_1$.
Via the standard relation between the mass $M$ and conformal dimension $\Delta$, we can obtain the value of conformal dimension of the dual operator as
\begin{align}
    \Delta_{\pm}=\frac{1}{2}\pm\s{\frac{1}{4}-M^2R_{dS}^2}=\frac{1}{2}\pm\frac{|p-2|}{2}
    \label{confdim(d=2)}
\end{align}
The analysis of the perturbation of the dS brane (\ref{perttoh}) predicts that the mass satisfy the upper bound (\ref{ubfgr}) and this guarantees that the conformal dimension $\Delta_{\pm}$ take only real values.

If we try to consider the other mass range $M^2R_{dS}^2>1/4$, the only possibility that the brane profile becomes real valued (we set $p=2+i\gamma$) is to choose
\begin{align}
    Z(t)\simeq -\ap t+\beta t^{2+i\gamma}+\beta^* t^{2-i\gamma}.
\end{align}
In this case, we can easily see that $\ddot{Z}(t)$ takes both positive and negative sign. In the latter case we find $\dot{\phi}^2< 0$, violating the null energy condition. Thus, this does not seem to be physically allowed solution in our setup embedded into the higher dimensional AdS.

At first, this result might look a little surprising from the conventional treatment of dS/CFT. This is because in dS/CFT, we normally allow the conformal dimensions to take complex values, which correspond to a scalar field with a large mass $M^2R_{dS}^2>1/4$. On the other hand, from our analysis in the light of AdS$_3/$CFT$_2$, we observed that the conformal dimensions should always take real values in order for the EOW brane to satisfy the null energy condition. 

However, this may be natural as the violation of null energy condition usually leads to that of unitarity and we indeed expect that the CFT dual to the dS gravity is non-unitary. 
Thus, our result suggests that if we restrict to the spectra of the dual CFT to be those have real valued conformal dimensions, we may be able to extract a unitary model of dS/CFT. This clearly deserves future studies.

When the profile approaches $Z(t)\sim\beta t^p$ as $t\to 0$, then its derivatives $\dot{Z}=\beta p t^{p-1}$ and $\ddot{Z}=\beta p(p-1)t^{p-2}$ imply that $p>1,\,\,\beta>0$, which tells us that this is type II.
Neumann boundary condition is 
\begin{align}
    \dot{\phi}^2
    % &=\frac{\ddot{Z}}{2Z\s{1-\dot{Z}^2}} \\
    =\f{\beta p(p-1)t^{p-2}}{2\beta t^p\sqrt{1-\beta^2p^2t^{2p-2}}} 
    \simeq \f{p(p-1)}{2}t^{-2},
\end{align}
thus
\begin{align}
    \phi-\phi_*\simeq \pm\sqrt{\f{p(p-1)}{2}}\log{t}.
\end{align}
The potential is 
\begin{align}
    V
    % &=\frac{- Z\ddot{Z}}{2(1-\dot{Z}^2)^{3/2}}+\frac{1}{\sqrt{1-\dot{Z}^2}} \\
    % &=\f{-\beta t^p\beta p(p-1)t^{p-2}}{2(1-\beta^2p^2t^{2p-2})^{3/2}}+\f{1}{\sqrt{1-\beta^2p^2t^{2p-2}}} \\
    % &\simeq -\f{\beta^2p(p-1)}{2}t^{2p-2}+1+\f{1}{2}\beta^2p^2t^{2p-2} \\
    % &\simeq1+\f{\beta^2 p}{2}t^{2p-2} \\
    \simeq 1+\f{\beta^2 p}{2}\exp\left[\pm2\sqrt{\f{2(p-1)}{p}}(\phi-\phi_*)\right].\label{eq:exponential_potential}
\end{align}  
This again shows that we need $V\geq 1$ in order for type II brane to reach the AdS$_3$ boundary.

\subsection{Examples of type I branes}
\subsubsection{Example(I-i): $Z(t)=\beta(t_0^2-t^2)$}
\label{Pointype1ex2}

\begin{figure}[htbp]
    \centering
    \includegraphics[width=0.6\linewidth]{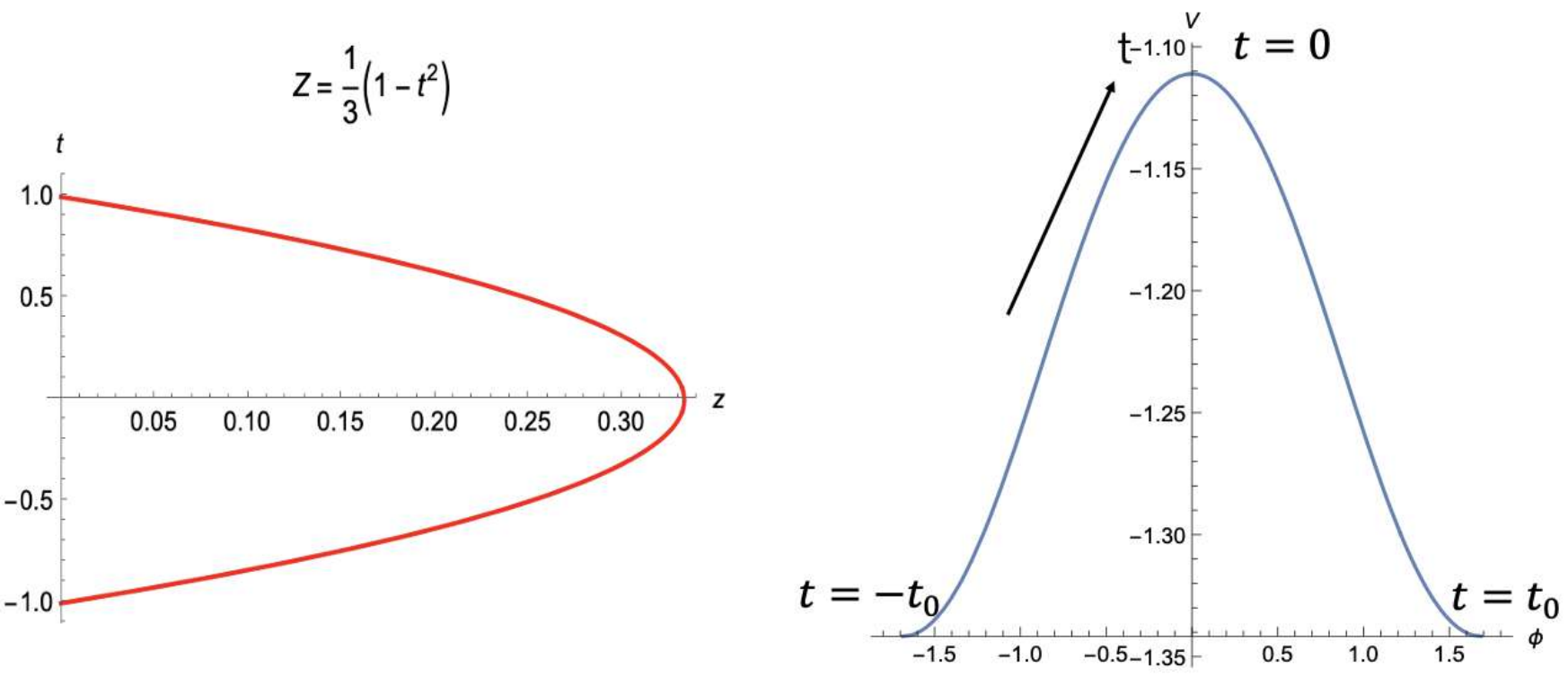}
    \caption{The left side is EOW brane setup. The right side is $V$-$\phi$ graph. We choose $\beta=\frac{1}{3}$ and $t_0=1$, so $Z(t)=\frac{1}{3} (1-t^2)$. }
    \label{fig:t^2}
\end{figure}

As an example we choose 
\ba
Z(t)=\beta(t_0^2-t^2),\quad -t_0 < t < t_0
\ea
This setup is showed in Fig.\ref{fig:t^2}.
Since the EOW brane should be time-like we require $2\beta t_0<1$.
The boundary condition is solved by choosing
\ba
&& \dot{\phi}^2=\frac{1}{(t_0^2-t^2)\s{1-4t^2\beta^2}},\label{twosola}\\
&& V(\phi)=-\frac{1+t_0^2\beta^2-5t_0^2\beta^2}{(1-4\beta^2 t^2)^{3/2}}.\label{twosolb}
\ea
Near the AdS boundary $z\to 0$ and $t\to t_0$, we can solve (\ref{twosola}) as
\ba
\phi\simeq \phi_*-\frac{\s{2(t_0-t)}}{\s{t_0}(1-4\beta^2 t_0^2)^{1/4}}.
\ea
The behavior of the potential is found from (\ref{twosolb})
\ba
V\simeq -\frac{1}{\s{1-4\beta^2 t_0^2}}+\frac{t_0^2\beta^2}{1-4t_0^2\beta^2}(\phi_*-\phi)^2.  \label{potone}
\ea
Since the induced metric near the tip $t=t_0$ reads
\ba
ds^2\simeq \frac{-(1-4t_0^2\beta^2)dt^2+dx^2}{4t_0^2\beta^2(t_0-t)^2},
\ea
We can read off the dS$_2$ radius $t=t_0$ as $R^2_{dS}=(1-4t_0^2\beta^2)/4t_0^2\beta^2$.
Thus the potential (\ref{potone}) looks like 
\ba
V(\phi)\simeq -\frac{1}{\s{1-4\beta^2 t_0^2}}+\frac{1}{4R^2_{dS}}(\phi_*-\phi)^2.
\ea
Since we find the mass of scalar field on dS$_2$ is $R^2_{dS}M^2=\frac{1}{4}$, we can obtain the conformal dimension of the dual operator via $\Delta=\frac{1}{2}+\s{\frac{1}{4}-M^2R_{dS}^2}=\frac{1}{2}$. $V$-$\phi$ graph is sketched in Fig.\ref{fig:t^2}.

\subsubsection{Example(I-ii): $Z(t)=\beta \cos \frac{t}{t_0}$}
\begin{figure}[H]
    \centering
    \includegraphics[width=0.6\linewidth]{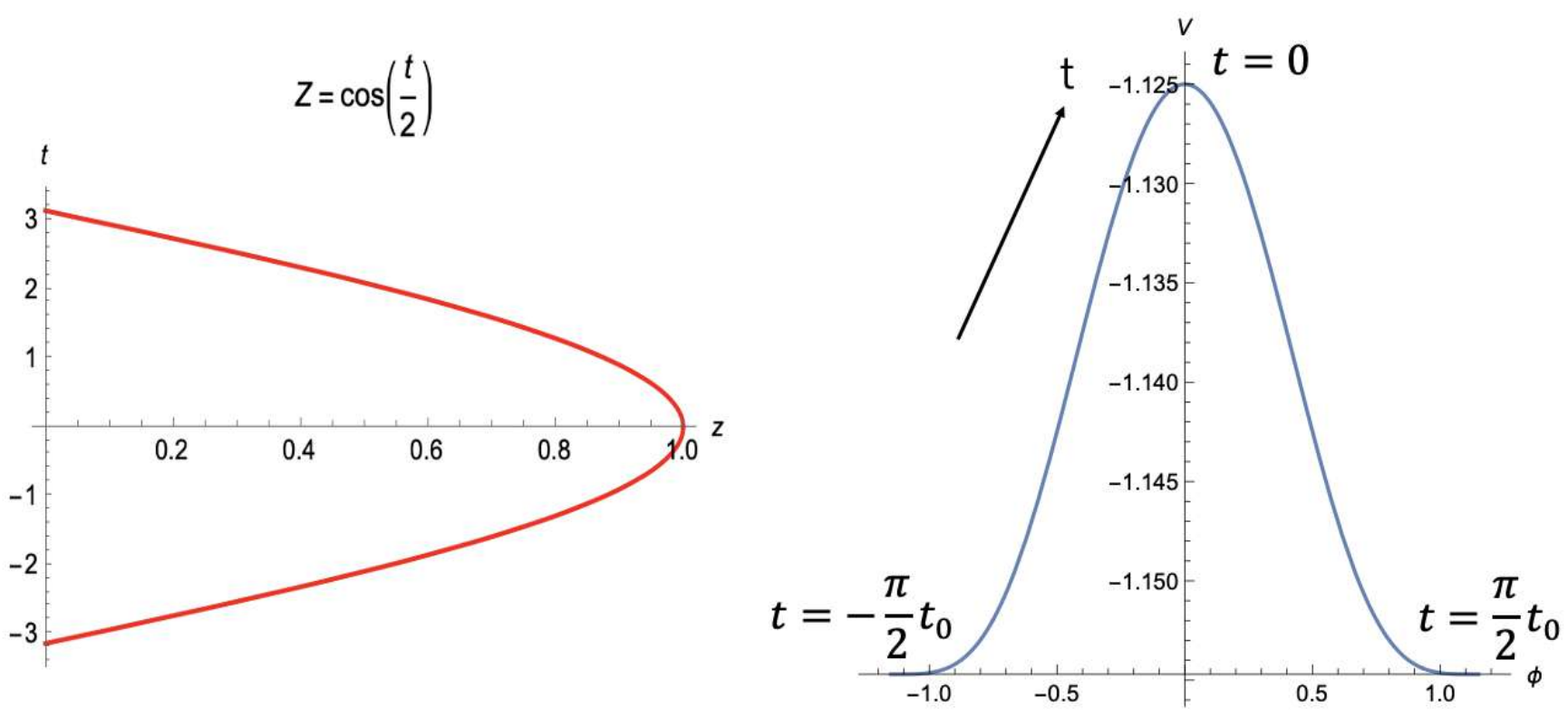}
    \caption{The left side is EOW brane setup. The right side is $V$-$\phi$ graph. We choose $\beta=1$ and $t_0=2$, so $Z(t)=\cos \frac{t}{2}$. }
    \label{fig:cos}
\end{figure}

As an example we choose 
\ba
Z(t)=\beta \cos \frac{t}{t_0},\quad -\frac{\pi}{2}t_0 < t < \frac{\pi}{2}t_0.
\ea
This setup is showed in Fig.\ref{fig:cos}.
Since the EOW brane should be time-like we require $1-\frac{\beta^2}{t_0^2}>0$.
The boundary condition is solved by choosing
\ba
&& \dot{\phi}^2=\frac{1}{2t_0^2\s{1-\frac{\beta^2}{t_0^2}\sin \frac{t}{t_0}}},\label{twosola4.1}\\
&& V(\phi)=\frac{-4+\frac{\beta^2}{t_0^2}-3\frac{\beta^2}{t_0^2}\cos{2\frac{t}{t_0}}}{\s{2}\left(2-\frac{\beta^2}{t_0^2}+\frac{\beta^2}{t_0^2}\cos{2\frac{t}{t_0}}\right)^{3/2}}.\label{twosolb4.2}
\ea
Near the AdS boundary $z\to 0$ and $t\to \frac{\pi}{2}t_0$, we can solve (\ref{twosola4.1}) as
\ba
\phi\simeq \phi_*-\frac{1}{\s{2}t_0\left(1-\frac{\beta^2}{t_0^2}\right)^{1/4}}\left(\frac{\pi}{2}t_0-t\right).
\ea
The behavior of the potential is found from (\ref{twosolb4.2})
\ba
V\simeq -\frac{1}{\s{1-\frac{\beta^2}{t_0^2}}}-\frac{3\frac{\beta^2}{t_0^2}}{2\left(1-\frac{\beta^2}{t_0^2}\right)^{3/2}}(\phi_*-\phi)^4.  \label{potone4}
\ea
Thus the scalar field is massless $M^2=0$.
We obtain the conformal dimension of the dual operator via $\Delta=\frac{1}{2}+\s{\frac{1}{4}-M^2R_{dS}^2}=1$. $V$-$\phi$ graph is sketched in Fig.\ref{fig:cos}.

\subsection{Examples of type II branes}

\subsubsection{Example(II-i): $Z(t)=\lambda\s{t^2+\ap^2}$}
\begin{figure}[H]
    \centering
    \includegraphics[width=0.6\linewidth]{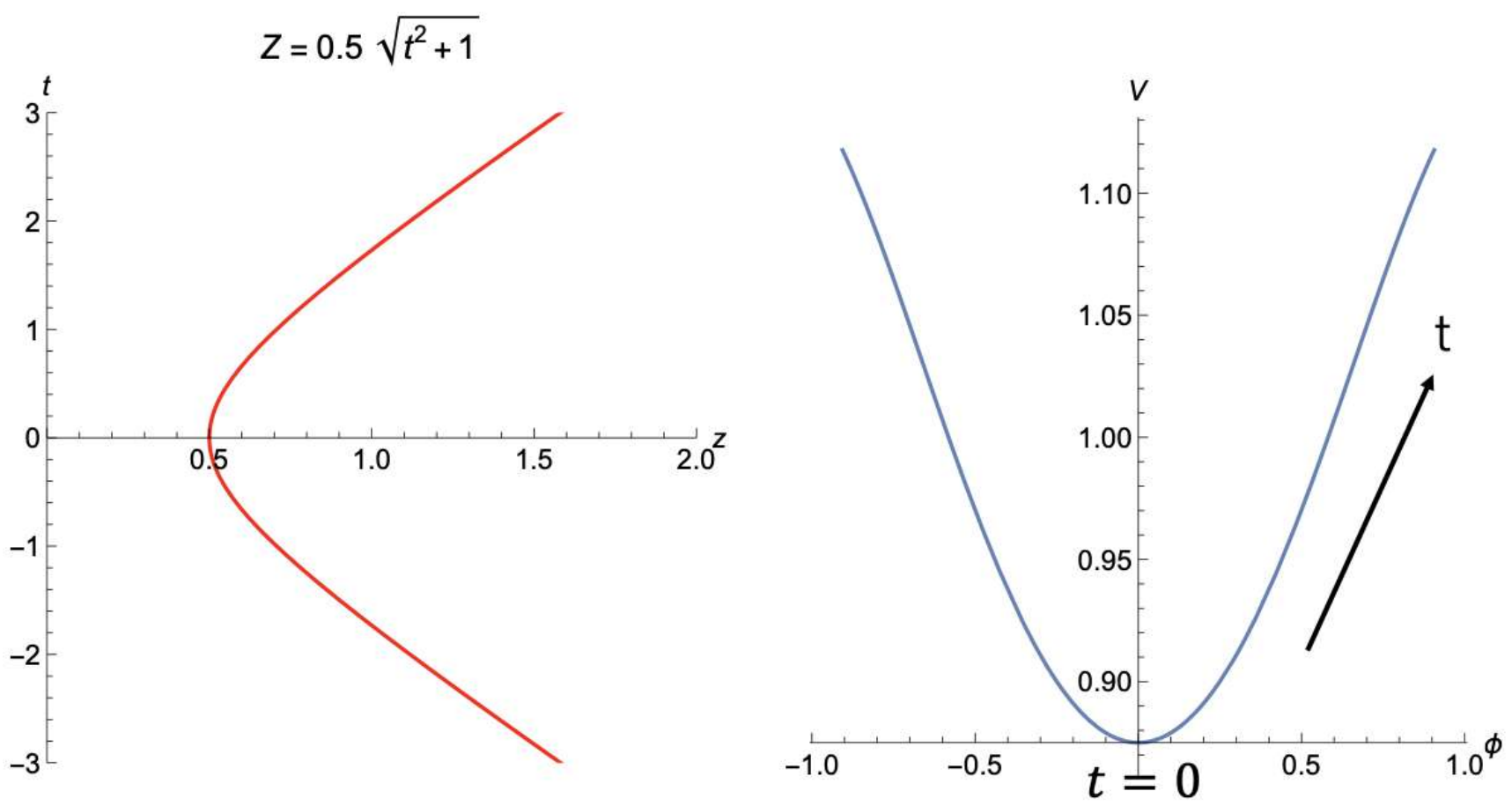}
    \caption{The left side is EOW brane setup. The right side is $V$-$\phi$ graph. We choose $\lambda=0.5$ and $\alpha=1$, so $Z(t)=0.5\sqrt{t^2+1}$. }
    \label{fig:root}
\end{figure}

We can choose the profile of EOW brane as 
\ba\label{eq:sqrt_Brane_in_poincare}
Z(t)=\lambda\s{t^2+\ap^2}.
\ea
The boundary conditions are solved by setting
\ba
&& \dot{\phi}^2=\frac{\ap^2}{2(\ap^2+t^2)^{3/2}\s{(1-\lambda^2)t^2+\ap^2}},\no
&& V(\phi)=\frac{\s{\ap^2+t^2}\left(2(1-\lambda^2)t^2+(2-\lambda^2)\ap^2\right)}{2\left((1-\lambda^2)t^2+\ap^2\right)^{3/2}}.
\ea
The profile and $V$-$\phi$ graph are in Fig.\ref{fig:root}.

\subsubsection{Example(II-ii): $Z(t)=a \log\left[\cosh\f{t}{t_0}\right]-b$}
\begin{figure}[H]
    \centering
    \includegraphics[width=0.6\linewidth]{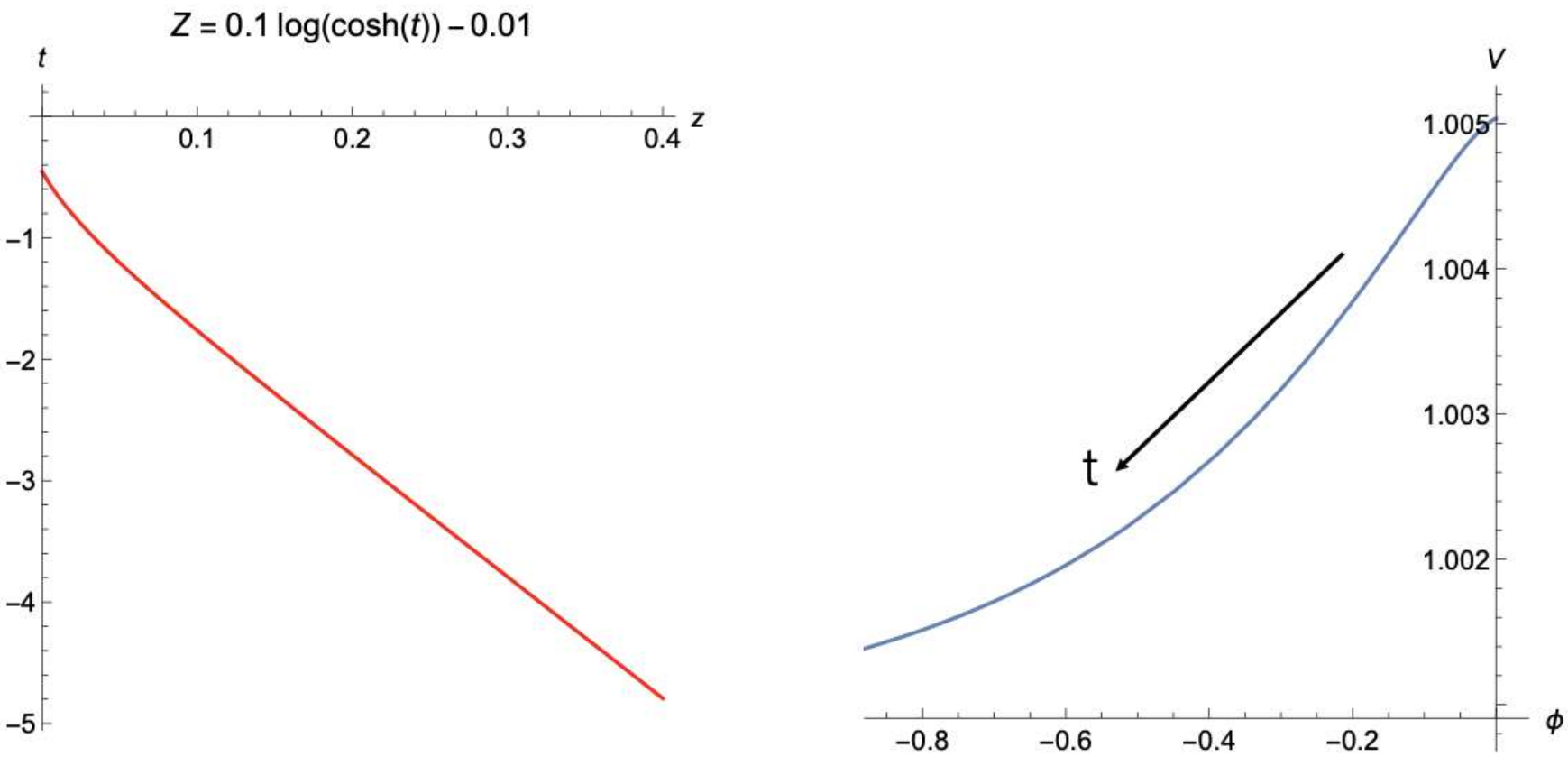}
    \caption{The left side is EOW brane setup. The right side is $V$-$\phi$ graph. We choose $a=0.1$, $b=0.01$ and $t_0=1$, so $Z(t)=0.1\log(\cosh (t))+0.01$. }
    \label{fig:logcosh}
\end{figure}

We can choose the profile of EOW brane as 
\be
Z(t)=a \log\left[\cosh\f{t}{t_0}\right] -b,\quad t < t_0\cosh^{-1}(e^{b/a})\label{logcosh}
\ee
The boundary conditions are solved by setting
\begin{align} 
\dot{\phi}^2 
&=\frac{a}{2 t_0 \cosh^2\left(\frac{t}{t_0}\right)\sqrt{t_0^2-a^2 \tanh ^2\left(\frac{t}{t_0}\right)} \left[a\log \left(\cosh \left(\frac{t}{t_0}\right)\right)-b\right]} \\
V(\phi)
&=-\frac{\f{at_0}{\cosh^2\left(\frac{t}{t_0}\right)}  \left[a\log \left(\cosh \left(\frac{t}{t_0}\right)\right)-b-2a\right]+2 t_0 (a^2-t_0^2) }{2 \left(t_0^2-a^2 \tanh^2\left(\frac{t}{t_0}\right)\right)^{3/2}}
\end{align}
 The profile and $V$-$\phi$ graph are in Fig.\ref{fig:logcosh}.

\subsection{Branes in Higher Dimensions}\label{sec:higherds}
In higher dimensional AdS/CFT, due to the back-reaction of the EOW brane, we cannot analytically find gravity solutions in general. However,  we can still find a special class of solutions by focusing on an EOW in the Poincar\'{e} AdS$_{d+1}$:
\ba
ds^2=\frac{dz^2-dt^2+dx_1^2+\cdots+dx_{d-1}^2}{z^2}.
\ea
We again specify the shape of the EOW brane $Q$ and the brane-localized scalar field on $Q$ by $z=Z(t)$ and $\phi=\phi(t)$. The induced metric on $Q$ reads
\ba
ds^2=\frac{-(1-\dot{Z}^2)dt^2+dx_1^2+\cdots+dx_{d-1}^2}{Z^2}.
\ea
The normal vector of the EOW brane is
\ba
(N^z,N^t,N^{x_1},\cdots,N^{x_{d-1}})=\frac{\epsilon Z}{\sqrt{1-\dot{Z}^2}}(1,-\dot{Z},0,\cdots,0),
\ea
where $\epsilon=\pm 1$. As before $\eps=1$ and $\eps=-1$ correspond to type I and type II, respectively. Then, the extrinsic curvature satisfies
\ba
K_{tt}-h_{tt}K=\epsilon \frac{-(d-1) \sqrt{1-\dot{Z}^2}}{Z^2},\\
K_{x_i x_i}-h_{x_i x_i}K=\epsilon \frac{(d-1)(1-\dot{Z}^2)-Z\ddot{Z}}{Z^2 \left(1-\dot{Z}^2\right)^{3/2}},
\ea
and the boundary condition (\ref{KEOM}) leads to
\ba
\dot{\phi}^2 &&= -\epsilon \frac{\ddot{Z}}{2Z \sqrt{1-\dot{Z}^2}},\\
V(t) &&=\epsilon \frac{-2(d-1)(1-\dot{Z}^2)+Z\ddot{Z}}{2(1-\dot{Z}^2)^{3/2}}.
\ea

We again consider a small perturbation around the dS$_{d}$ brane solution 
$Z=-\ap t$ which is assumed to be the following form:
\ba
Z(t)\simeq -\ap t-\epsilon \beta t^p+ o(t^p),
\ea
in the $t\to 0$ limit. Since it is time-like and satisfies the null energy condition we require
\ba
0<\ap<1,\ \ \ \beta>0,\ \ \  p>1.
\ea
Then we obtain the following behaviors
\ba
&& \dot{\phi}^2\simeq \frac{p(p-1)\beta}{2\ap\s{1-\ap^2}}t^{p-3},\no
&& V(\phi)\simeq -\epsilon \frac{d-1}{\s{1-\ap^2}}-\frac{p(p-2d+1)\ap\beta}{2(1-\ap^2)^{3/2}}t^{p-1},
\ea
By solving the first equation we find
\ba
\phi\simeq \phi_*\pm \s{\frac{2p\beta}{(p-1)\ap\s{1-\ap^2}}}t^{\frac{p-1}{2}}.
\ea
Thus the potential can be rewritten as 
\ba\label{eq:potential_V_phi_of_general_arg}
V(\phi)\simeq -\e\frac{(d-1)}{\s{1-\ap^2}}+\frac{\ap^2(p-1)(2d-1-p)}{4(1-\ap^2)}(\phi-\phi_*)^2+\ddd.
\ea
Since the dS radius in this case at $t=0$ reads $R_{dS}^2=\frac{1-\ap^2}{\ap^2}$, we find
\ba
M^2 R_{dS}^2=\frac{(p-1)(2d-1-p)}{4}\leq \frac{(d-1)^2}{4}.
\ea
The dual conformal dimension reads 
\ba
\Delta=\frac{d-1}{2}+\frac{|d-p|}{2}.
\ea
This result is consistent with the $d=2$ case (\ref{confdim(d=2)}).
We obtain the implication to dS$_{d}/$CFT$_{d-1}$ in the same way as the one we already discussed for $d=2$.

%%%%%%%%%%%%%%%%%%%%%%%%%%%%%%%%%%%%%%%%%%%%%%%%%%%%%%%%
%%%%%%%%%%%%%%%%%%%%%%%%%%%%%%%%%%%%%%%%%%%%%%%%%%%%%%%%
\section{Holographic time-like g-theorem}
\label{sec:g-th}
%%%%%%%%%%%%%%%%%%%%%%%%%%%%%%%%%%%%%%%%%%%%%%%%%%%%%%%%
%%%%%%%%%%%%%%%%%%%%%%%%%%%%%%%%%%%%%%%%%%%%%%%%%%%%%%%%

In the absence of the scalar field, the solution of EOW brane is given by a dS$_2$ branes. Its CFT dual is given by a BCFT in the type I case and by a CFT coupled to dS$_2$ gravity in the type II case, as depicted in Fig.\ref{fig:hyperplane}. Therefore we can regard perturbations of the brane solutions by turning on the scalar field $\phi$ can be regarded as those of the dual CFT localized along the line at $t=0$. Thus this can be regarded as a boundary perturbation in a BCFT in the type I case and as an interface perturbation in the system of the CFT coupled to gravity in the type II case.  

This motivates us to consider a possibility of monotonicity under the renormalization group (RG) flow trigger by such a relevant perturbation localized on a space-time line. In the standard BCFT with a time-like boundary, such a monotonicity is known as g-theorem \cite{Affleck:1991tk,Friedan:2003yc}. We can reformulate the g-theorem in terms of entanglement entropy \cite{Harper:2024aku}. 
Consider a two dimensional BCFT on the right half plane $x>0$. We choose the subsystem $A$ to be the interval $0\leq x\leq l$ at the time slice $t=0$. 
As found in \cite{Calabrese:2004eu}, the entanglement entropy $S_A=-\mbox{Tr}[\rho_A\log\rho_A]$ defined for the reduced density matrix $\rho_A$ takes the following form 
\ba
S_{A} &&= \frac{c}{6}\log\frac{l}{\ep_{\rm{UV}}} + \log g,
\label{g-function}
\ea
where $\ep_{\rm{UV}}$ is the UV cut off and $c$ is the central charge. In this expression, the constant $g$ coincides with the $g$ function and $\log g$ is also called the boundary entropy. If we perform a boundary relevant perturbation, $g$ starts to depend on the subsystem size $l$. The g-theorem argues that $g$ is monotonically decreasing as a function of $l$ \cite{Harper:2024aku}. Below we would like to explore a similar theorem in our setup with the space-like boundary (type I) or interface (type II). 

For this we will employ the holographic calculation of entanglement entropy
\cite{Ryu:2006bv,Ryu:2006ef,Hubeny:2007xt} in the presence of EOW brane \cite{Takayanagi:2011zk}. The entanglement entropy can be computed in AdS$_3/$BCFT$_2$ by
\ba
S_{A} =\frac{L(\Gamma_A)}{4 G_{\rm{N}}} = \frac{c}{6}L(\G_A),
\label{RTformula}
\ea
where $\Gamma_A$ is the space-like geodesic which ends on the boundary of $A$ and $L(\Gamma_A)$ is its geodesic length. A new special feature is that $\Gamma_A$ is allowed to end also on the EOW brane and we need to extremize the geodesic length with respect to the end point on the brane.

We will choose the subsystem $A$ to be a semi-infinite interval $0\leq x<\infty$ at a fixed time $t=\tilde{t}(<0)$. We now pick up the finite term in the entanglement entropy $S_A$ and regard it as a function of the time $\tilde{t}$, which will be called a time-like g-function. Since $\tilde{t}$ is the only scale in the BCFT, it is natural to regard it as the RG scale when we perturb the BCFT. Indeed, below we will be able to confirm that the time-like g-function is indeed monotonically decreasing function of $|\tilde{t}|$ for both the type I and type II brane. We would also like to note that a closely related monotonic behavior, called $p$-theorem, for cross cap states was shown in the interesting paper \cite{Wei:2024zez}, by using holography. Though the setup is similar, we will show this monotonicity by using the holographic entanglement entropy.

It is intriguing to note that we can also interpret this entanglement entropy as the time-like entanglement entropy \cite{Doi:2022iyj,Doi:2023zaf,Narayan:2022afv} for which the subsystem is chosen to be the time-like interval $\tilde{t}\leq t \leq 0$ at a fixed value of $x$. This is because the holographic entanglement entropy is given by the same geodesic which connects $(t,x)=(\tilde{t},0)$ to a point on the EOW brane. This interpretation is very similar to the original g-theorem setup because the $g$ function is a monotonic function of the length of interval i.e. $|\tilde{t}|$.

\subsection{g-theorem in the type I case}

\begin{figure}[htbp]
    \centering
    \includegraphics[width=0.35\linewidth]{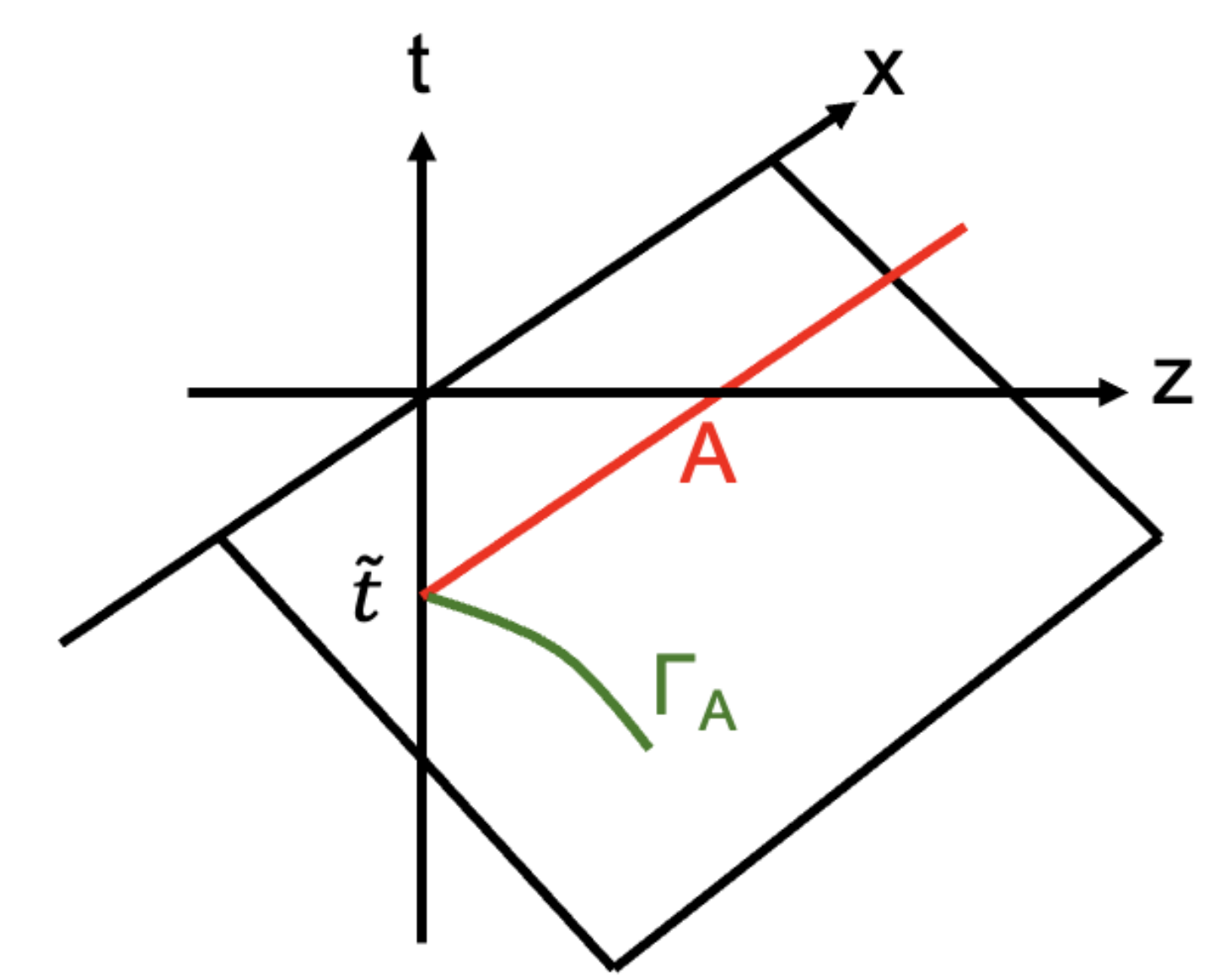}
    \includegraphics[width=0.28\linewidth]{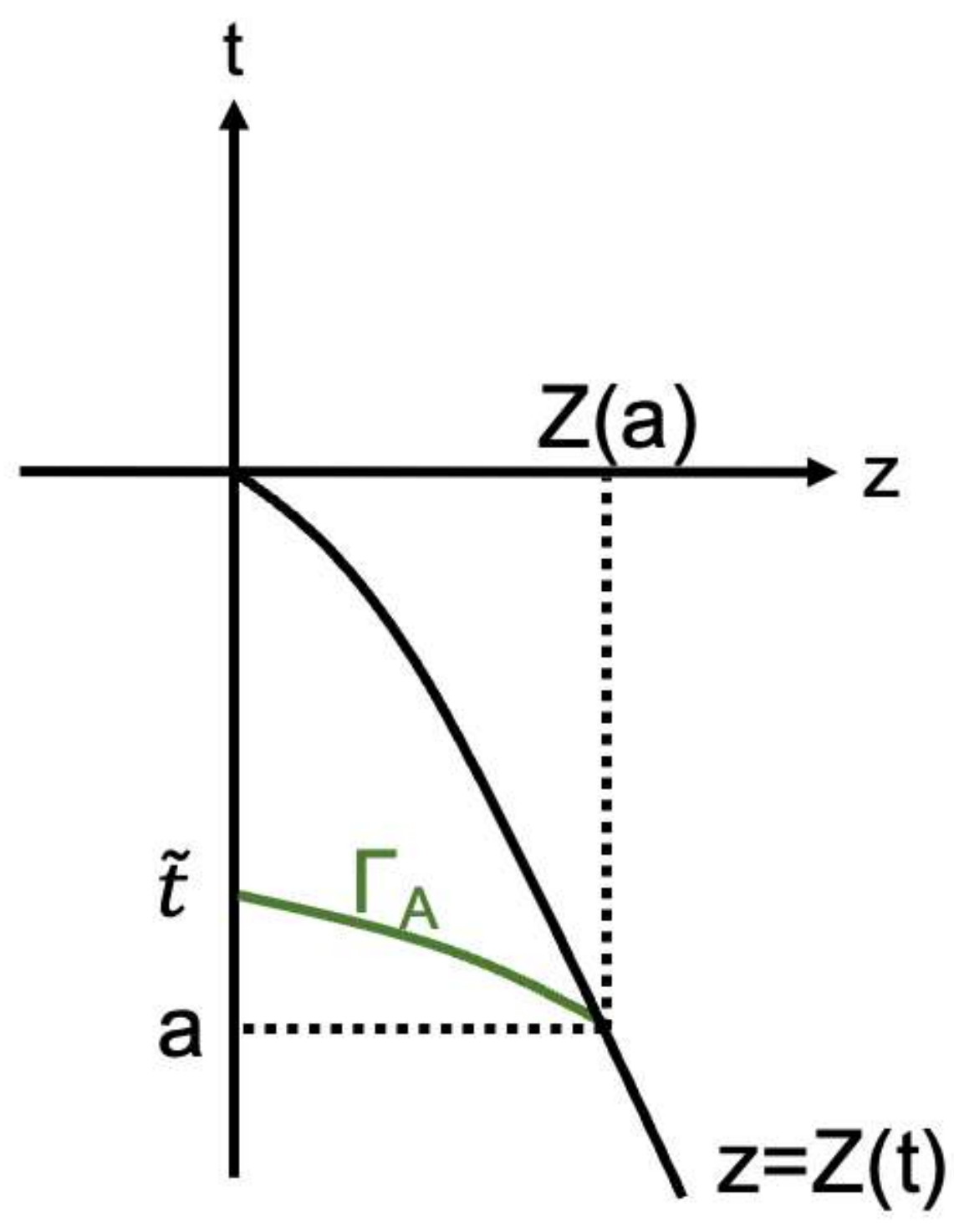}
    \caption{The profile of the geodesic $\Gamma_A$ which computes the entanglement entropy (left) and the calculation of its geodesic length (right). For type I brane, time coordinates $a$ of the endpoint on the EOW brane is negative.}
    \label{fig:type1hyp}
\end{figure}

First we consider the dS$_2$ EOW brane $Z(t)=-\lambda t$ before we turn on the scalar field. As we mentioned we choose the subsystem $A$ to be the semi-infinite line $x>0$ at a fixed time $t=\tilde{t}$. Then $S_A$ can be found from the geodesic which connects $(t,z)=(\tilde{t},\ep_{\rm{UV}})$ and $(t,z)=(a,z(a))$ as depicted in the left panel of Fig.\ref{fig:type1hyp}. By maximizing the geodesic length as a function of $a$, we find
\ba
a=\frac{\tilde{t}}{\sqrt{1-\lambda^2}}.
\ea
Thus, we obtain the following result of holographic entanglement entropy
(\ref{RTformula}):
\ba
    S_{A}=\frac{c}{6}\log\frac{|\tilde{t}|}{\ep_{\rm{UV}}}+\frac{c}{6} \log \left(\frac{2(1-\sqrt{1-\lambda^2})}{\lambda}\right).
\ea
The second term in this equation gives $\log g$, namely the time-like $g$ function.

Now we extend this calculation to the general EOW brane profile $z=Z(t)$, in the presence of non-trivial scalar field $\phi(t)$. We can again compute $S_A$ from the geodesic length (see the right panel of Fig.\ref{fig:type1hyp}).  The length of $\Gamma_A$
reads
\ba\label{eq:type_I_gamma_A}
L(\Gamma_A) \simeq \log\left[\frac{Z(a)^2-(a-\tilde{t})^2}{Z(a)\ep_{\rm{UV}}}\right],
\ea
in the limit $\ep\to 0$. We choose $a$ which maximizes the geodesic length $L(\Gamma_{\rm{A}})$, leading to the relation between $a$ and $\tilde{t}$ as follows:
\ba
a-\tilde{t}=\frac{Z(a)}{\dot{Z}(a)}(1-\sqrt{1-{\dot{Z}(a)}^2}).
\ea
Thus, we can calculate the derivative of $\log g$ with respect to $|\tilde{t}|$.
\ba
\frac{6}{c}|\tilde{t}|\frac{d \log g}{d |\tilde{t}|}
=|\tilde{t}|\frac{d L(\Gamma_A)}{d|\tilde{t}|}-1 
=\frac{a \dot{Z}(a)-Z(a)}{Z(a)\sqrt{1-\dot{Z}(a)^2}} 
\ea

We can find that the numerator $a \dot{Z}(a)-Z(a)$ vanishes in the limit $a\to 0$ and that the derivative of the numerator, $\frac{d}{da}(a \dot{Z}(a)-Z(a))=a \ddot{Z}(a)$, is positive owing to the type I null energy condition $\ddot{Z}(a)\leq 0$. This shows $a \dot{Z}(a)-Z(a)\leq 0$ for $a\leq 0$. Thus, finally we can conclude that 
\ba
\frac{d\log g}{d |\tilde{t}|}\leq 0,
\ea
which gives the holographic proof of the time-like $g$-theorem.

%%%%%%%%%%%%%%%%%%%%%%%%%%%%%%%%%%%%%%
\subsection{g-theorem in the type II case}
Now we would like to turn to the type II brane and would like to evaluate $S_A$ for the same subsystem $A$. At first, one may be puzzled because the computation of holographic entanglement entropy (\ref{RTformula}) is not straightforward because there is no space-like geodesic which ends on $(t,z)=(\tilde{t},\ep_{\rm{UV}})$ and on a point on the EOW brane. 

\begin{figure}[htbp]
        \centering
        \includegraphics[width=0.5\linewidth]{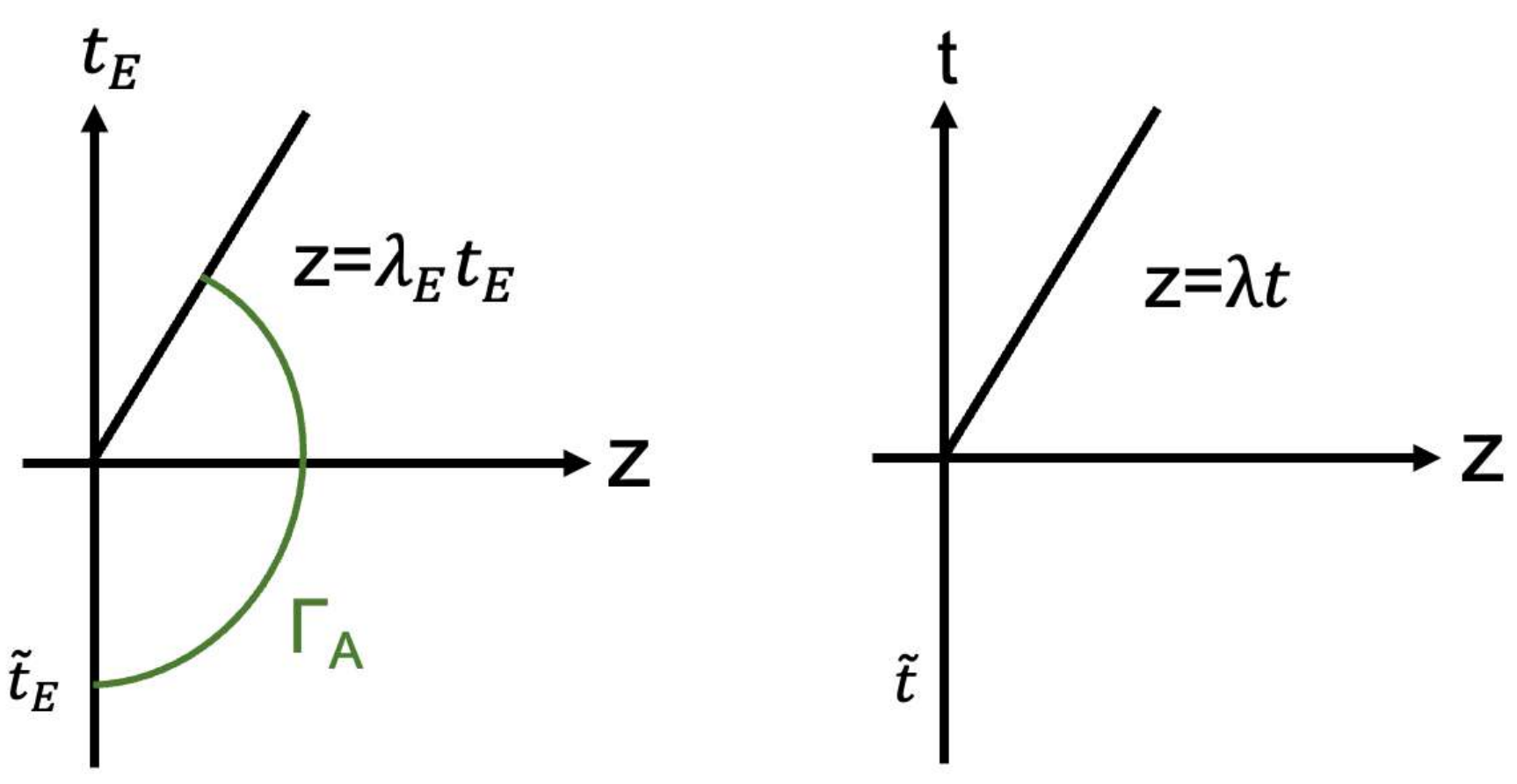}
        \caption{Type II Hyperplane in the Euclidean(left) and Lorentzian(right) setup}
        \label{fig:EuLosetup}
\end{figure}

We resolve this problem in two ways as we will explain below.
First, we consider the type II dS$_2$ brane in the Euclidean setup described in the left panel of Fig.\ref{fig:EuLosetup}. We obtain the entanglement entropy 
\ba
S^{(E)}_A=\frac{c}{6} \log \left[-\frac{2\tilde{t}_E}{\ep_{\rm{UV}}}\left(\frac{1+\sqrt{1+\lambda_E^2}}{\lambda_E}\right)\right].
\ea
Now we perform the Wick rotation $t_E=i t$ and introduce $\lambda=i \lambda_E$ (i.e. the right panel of Fig.\ref{fig:EuLosetup}). This leads to the entanglement entropy in our Lorentzian setup as follows:
\begin{align}
S^{(L)}_{A}&=\frac{c}{6} \log \left[\frac{2 \tilde{t}}{\ep_{\rm{UV}}}\left(\frac{1+\sqrt{1-\lambda^2}}{\lambda}\right)\right]\\
&=\frac{c}{6}\log\frac{\abs{\tilde{t}}}{\e_{\rm{UV}}}+\frac{c}{6}\log\left[2\frac{1+\sqrt{1-\l^2}}{\l}\right]+\frac{c}{6}\pi i.
\label{Lotype2}    
\end{align}

Now we move on to the general profile of the type II EOW brane. It is difficult to consider the Euclidean counterpart. Rather, it is convenient to calculate the entanglement entropy directly via \eqref{eq:type_I_gamma_A}.

We take a subregion $A$ in the same way as Fig.\ref{fig:type1hyp}. The EOW brane is supposed to be characterized by $z=Z(t)$, whose argument $t$ is positive since the brane is type II. As we mentioned, any space-like geodesic starting at $(t,z)=(\tilde{t},\e_{\rm{UV}})$ cannot end on the brane directly, which is because the argument of logarithm in \eqref{eq:type_I_gamma_A} must be negative in type II case. This is led by the fact $\dot{Z}(t)\leq 1$, which says immediately
\begin{equation}
    Z(a)\leq a\leq a - \tilde{t}.
\end{equation}
Note that $a$ is positive but $\tilde{t}$ is negative.

Since the argument of logarithm function is negative, the entanglement entropy will have the imaginary part:
\begin{equation}\label{eq:type_II_gen_ent}
    S_{A}(\tilde{t}) = \frac{c}{6}\log\left[\frac{-Z(a)^2+(a-\tilde{t})^2}{Z(a)\ep_{\rm{UV}}}\right] + \frac{c}{6}\pi i,
\end{equation}
where $a$ is chosen so that the real part of the entanglement entropy is minimized.

Now we define a boundary entropy for the type II brane as
\begin{equation}
    \log g = S_A(\tilde{t}) - \frac{c}{6}\log\frac{\abs{\tilde{t}}}{\e_{\rm{UV}}} - \frac{c}{6}\pi i = \frac{c}{6}\log\left[\frac{-Z(a)^2 + (a-\tilde{t})^2}{2Z(a)\abs{\tilde{t}}}\right].
\end{equation}
The derivative of the boundary entropy will be calculated as
\begin{equation}
    \frac{6}{c}\abs{\tilde{t}}\frac{d \log g}{d \abs{\tilde{t}}} = -\frac{a\dot{Z}(a)-Z(a)}{Z(a)\sqrt{1-\dot{Z}(a)^2}}.
\end{equation}
By similar discussion of the type I case, since $\ddot{Z}\geq 0$ according to the null energy condition for type II branes, we can conclude
\begin{equation}
    \frac{d}{d\abs{\tilde{t}}} \log g\leq 0,
\end{equation}
which is the time-like $g$-theorem for the type II brane.

\begin{figure}[htbp]
    \centering
    \includegraphics[width=0.30\linewidth]{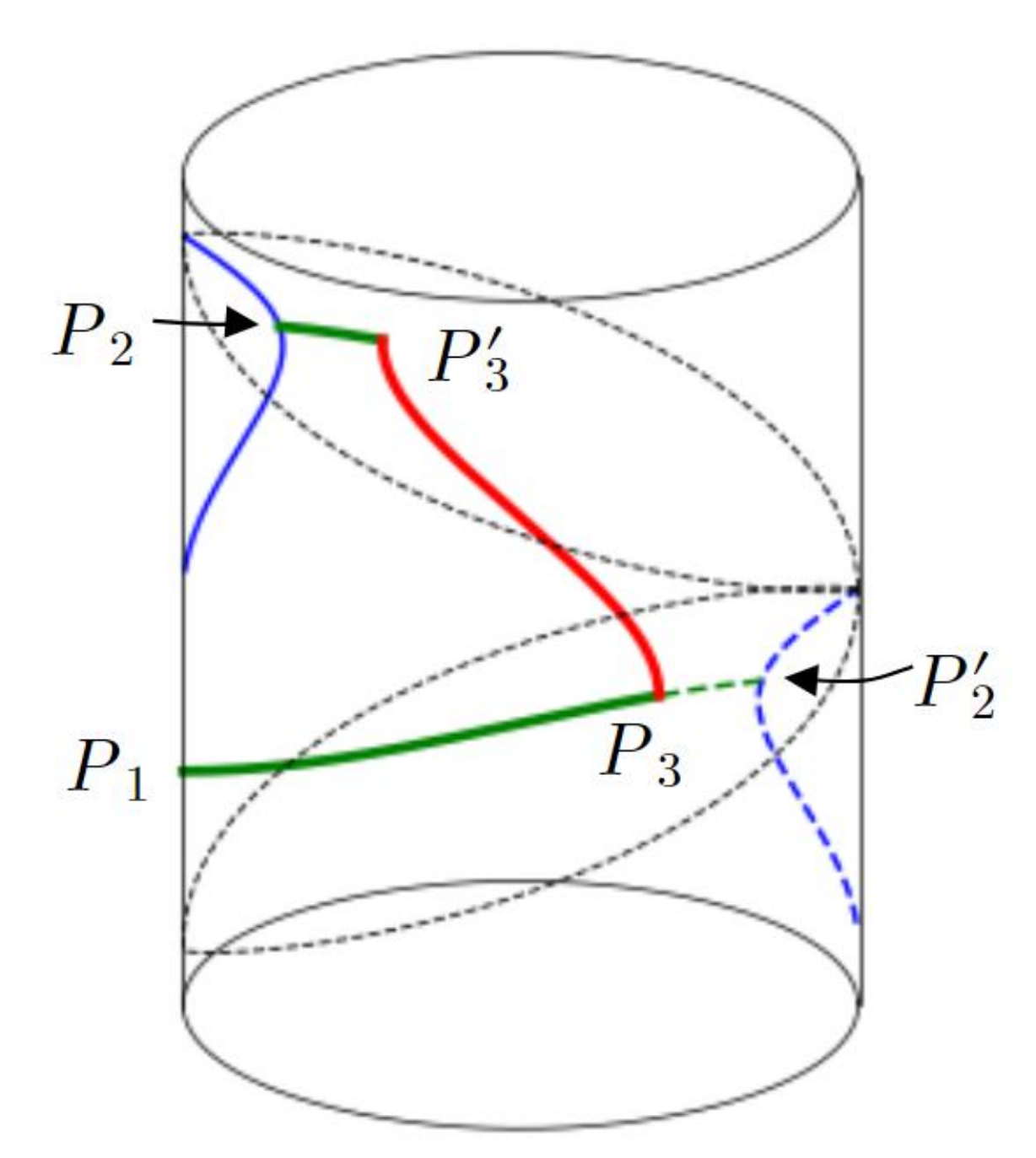}
    \hspace{0.05\linewidth}
 \includegraphics[width=0.25\linewidth]{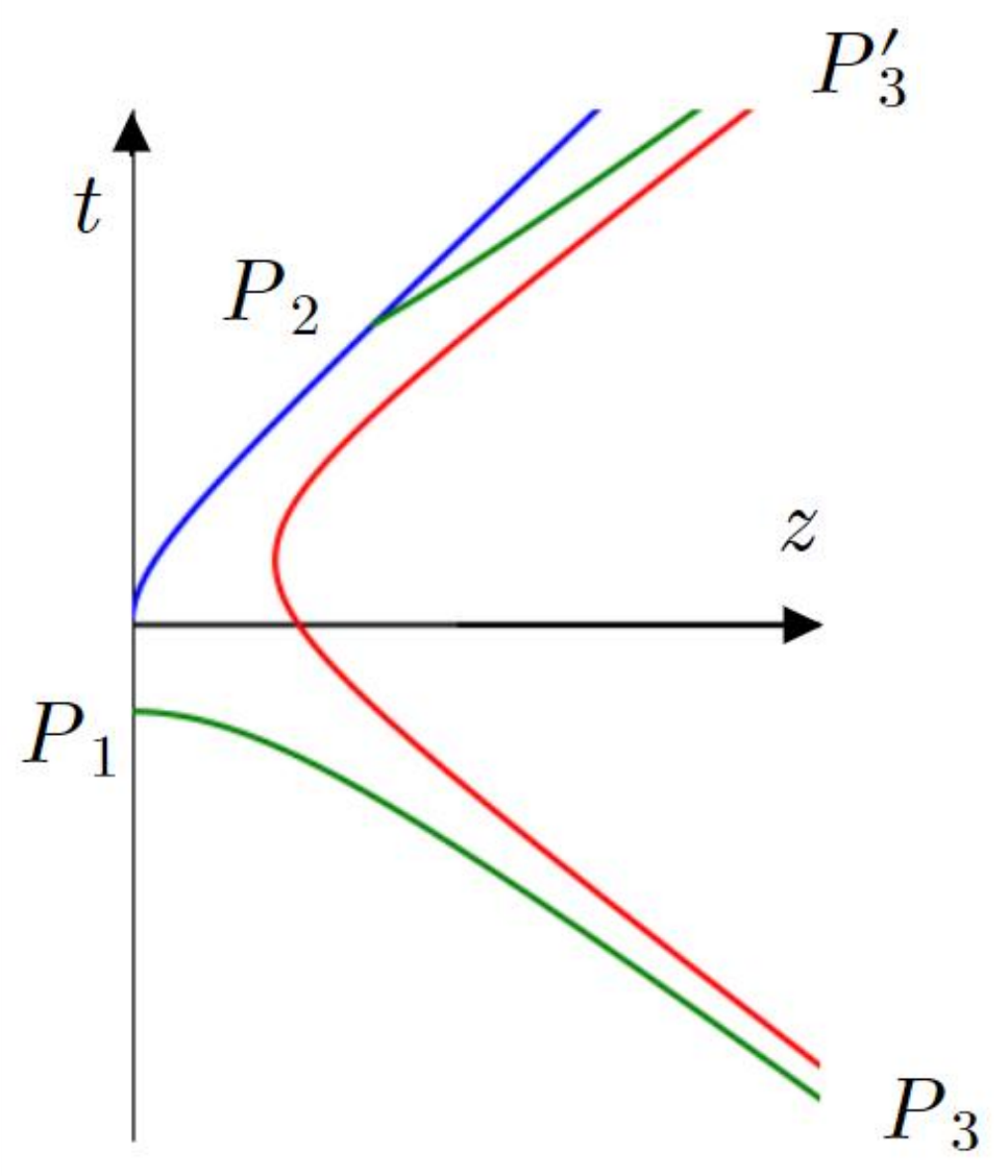}
    \hspace{0.05\linewidth}
 \includegraphics[width=0.30\linewidth]{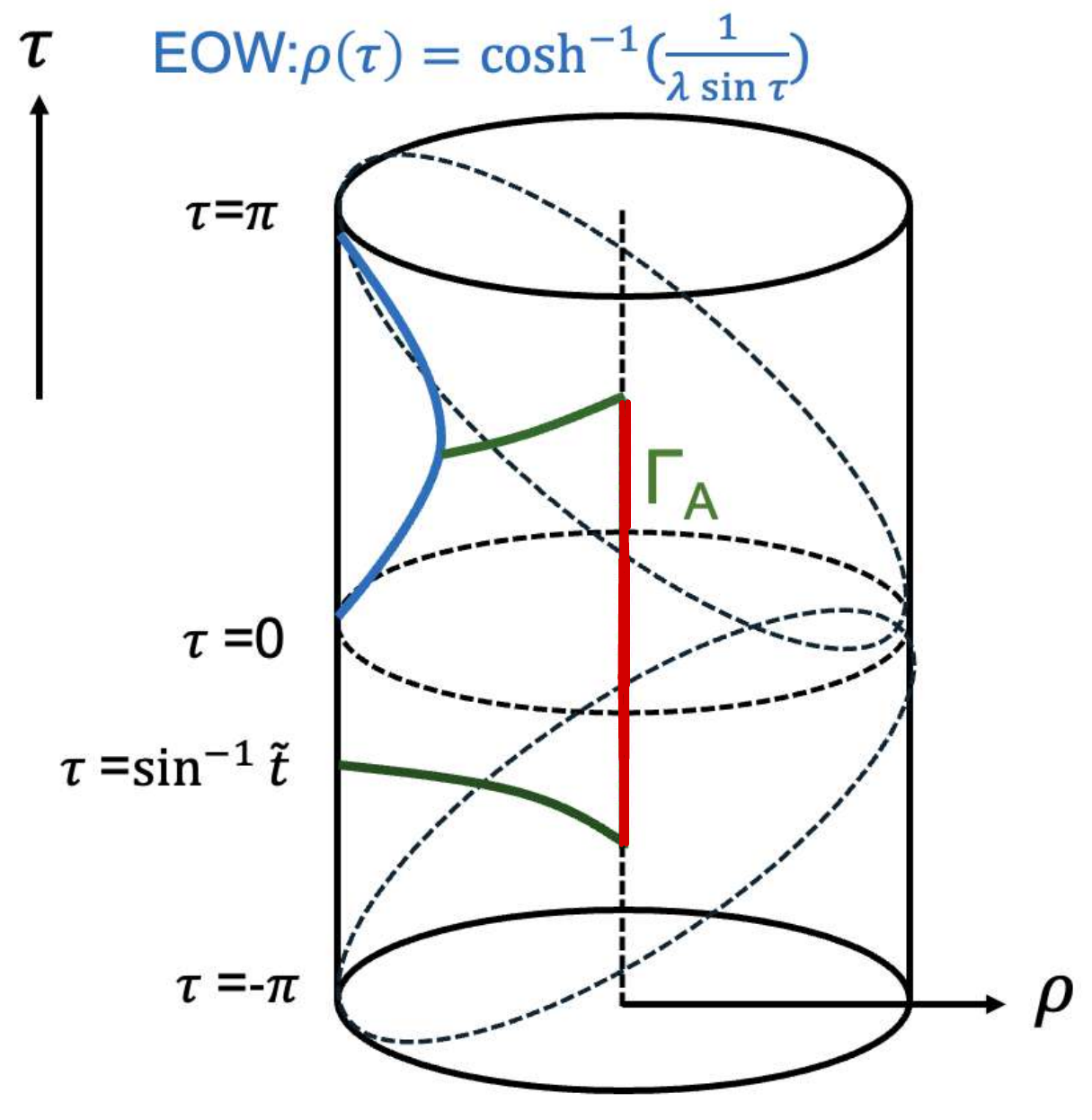}
    \caption{The left panel illustrates the profile of the brane and the RT surface in global AdS using \eqref{eq:embedding_poincare}, while the middle panel shows their counterparts in \Poincare AdS. The blue solid line represents the original EOW brane, while the blue dotted line represents its copy. Additionally, the green solid curves represent the space-like geodesics, which contribute to the real part of the entanglement entropy, and the red curve represents the time-like geodesic, which gives to the imaginary part. The right panel shows the configuration if the EOW brane is hyperplane $Z(t)=\l t$.}
    \label{fig:geodesics_of_type_II_EE}
\end{figure}

Now we present the geometric interpretation of the type II entanglement entropy in   \eqref{eq:type_II_gen_ent}. To do this, it is useful to embed the  \Poincare AdS into the global AdS $ds^2 = d\r^2 - \cosh^2\r d\t^2 + \sinh^2\r d\phi^2$ via
\begin{equation}
    \cosh\rho = \frac{\sqrt{t^{2} +D^{2}}}{z},\quad \tan \t = \frac{t}{D},\quad \tan\phi = \frac{x}{D - 1},\label{eq:embedding_poincare}
\end{equation}
where $D=(z^{2} +x^{2} -t^{2} +1)/2$. The left panel of Fig.\ref{fig:geodesics_of_type_II_EE} illustrates the embedding described above.

In Fig.\ref{fig:geodesics_of_type_II_EE}, the blue solid line represents the EOW brane and a point $P_1$ is the endpoint of the half-line $A$. Additionally, we place a copy of the EOW brane under the \Poincare patch, represented by the blue dotted curve. From the perspective of global AdS, we can calculate the geodesic length between the points $P_1$ and $P\pr_2$, which lies on the copy of the brane. In fact, this length is given by
\begin{equation}
    L(P_1P\pr_2) = \log\left[\frac{-Z(a)^2+(a-\tilde{t})^2}{Z(a)\ep}\right],
\end{equation}
which is exactly the same as the real part of the entanglement entropy. The point $P\pr_2$ is chosen so that the distance between $P_1$ and $P\pr_2$ is minimized.

Let the point $P_3$ be the intersection of the lower \Poincare horizon and the geodesic curve connecting $P_1$ and $P\pr_2$ and let the point $P\pr_3$ be the copy of $P_3$ on the upper \Poincare horizon. Note that the curve connecting $P_3$ and $P\pr_3$ is time-like and its length is given by $L(P_3P\pr_3)=\pi$, regardless of the choice of the brane as we will see soon. Since $L(P_1P\pr_2)=L(P_1P_3)+L(P\pr_3P_2)$, the entanglement entropy \eqref{eq:type_II_gen_ent} is identified as
\begin{equation}
    S_{A} = \frac{c}{6}\left(L(P_1P_3)+iL(P_3P\pr_3)+L(P\pr_3P_2)\right).
\end{equation}

In fact, for the hyperplane case $Z(t)=\l t$ in the right panel of Fig.\ref{fig:geodesics_of_type_II_EE}, we find that the length $L(P_1P_3) + L(P\pr_3P_2)$ coincides with the real part of \eqref{Lotype2}. Although the curve $P_3P\pr_3$ in the left panel of Fig.\ref{fig:geodesics_of_type_II_EE} is non-trivially curved, it can be straightened, like the right panel of Fig.\ref{fig:geodesics_of_type_II_EE}, using a different embedding than \eqref{eq:embedding_poincare} due to the isometry of AdS space. Consequently, we conclude that $L(P_3P\pr_3) = \pi$ for any brane configuration, which yields the imaginary contribution to the entanglement entropy \eqref{Lotype2}.

This complex-valued entropy has the same imaginary part as that of the time-like entanglement entropy \cite{Doi:2022iyj,Doi:2023zaf}. In fact, as we mentioned above, we can interpret $S_A$ as the time-like entanglement entropy for the time-like interval $\tilde{t}\leq t\leq 0$ at $x=0$.

%%%%%%%%%%%%%%%%%%%%%%%%%%%%%%%%%%%%%%%%%%%%%%%%%%%%%%%%
%%%%%%%%%%%%%%%%%%%%%%%%%%%%%%%%%%%%%%%%%%%%%%%%%%%%%%%%
\section{Liouville gravity description}
\label{sec:Liouville}

In this section, we will show that the dynamics on the EOW brane $Q$ can be approximated by a  Liouville gravity \cite{Polyakov:1981rd,Takayanagi:2018pml,Boruch:2020wax,Suzuki:2022xwv,neuenfeld2024Liouville} coupled to  scalar field when the EOW brane varies slowly, i.e. $\abs{\dot{Z}}\ll 1$.
This helps us to understand how the two-dimensional quantum gravity on the EOW brane looks like, which expected to be dual to the three dimensional classical gravity on the AdS$_3$ spacetime surrounded by the brane. For simplicity, we assume the brane has a translational invariance in the space $x$ direction and we take into account only the time-dependence.

\subsection{Analysis of brane action}

First, we focus on the type I brane. Given that the brane is determined by $z = Z(t)$, the induced metric on the brane is
\begin{equation}
    ds^2_Q = \frac{-(1-\dot{Z}^2)dt^2 + dx^2}{Z^2} = \frac{-d\t^2 + dx^2}{Z^2}.
\end{equation}
Here, we introduce a new coordinate $\t$, which is defined through
\begin{equation}
    \frac{d\t}{dt} = \sqrt{1 - \dot{Z}^2} = \frac{1}{\sqrt{1 + Z\prsq}},
\end{equation}
where the differential with respect to new coordinate $\t$ is denoted as $Z\pr$.

The total action, including the asymptotic surface $\Sigma$ at $z = \e_{\rm{UV}}$, is given by $I = I_M + I_\Sigma + I_Q$,
\begin{align}
    I_M &= \frac{1}{16\pi G_N}\int_M d^3x\sqrt{-g}(R-2\L) = \frac{L_x}{8\pi G_N}\int d\t\frac{\sqrt{1+Z\prsq}}{Z^{2}}-\frac{L_x L_t}{8\pi G_N\e^2_{\rm{UV}}},\\
    I_\Sigma &=\frac{1}{8\pi G_N}\int_\S dxdt\sqrt{-h}K=\frac{L_xL_t}{4\pi G_N\e^2_{\rm{UV}}},\\
    I_Q &=\frac{1}{8\pi G_N}\int_Q dxdt\sqrt{-h}(K - h^{tt}\dot{\phi}^2-V(\phi))\\
    &=\frac{L_x}{8\pi G_N}\int d\t\frac{-2-2Z\prsq +ZZ\prpr + Z^2\phi\prsq -\sqrt{1 + Z\prsq}V(\phi) }{Z^{2}\sqrt{1+Z\prsq}}.
\end{align}
Here we define $L_x$ and $L_t$ as infinite lengths in the $x$ direction and the $t$ direction and use $R=-6$ and $\L = -1$. Furthermore, by introducing $\varphi(\t):=-\log Z(\t)$, i.e. $ds^2_Q = e^{2\varphi}(-d\t^2 +dx^2)$, the total action is reduced to
\begin{equation}
    I = \frac{1}{8\pi G_N}\int d\t dx \left(\frac{-\varphi\prpr -e^{2\varphi}}{\sqrt{1+e^{-2\varphi }\varphi\prsq}}+\frac{\phi\prsq}{\sqrt{1+e^{-2\varphi }\varphi\prsq}}-e^{2\varphi}V(\phi)\right).
\end{equation}
Here we omit the constant term. Although this integral contains the second derivative, it is eliminated by a partial integral as follows:
\begin{align}
    I &= \frac{1}{8\pi G_N}\int d\t dx\left[e^{\varphi }\varphi\pr\sinh^{-1}( e^{-\varphi }\varphi\pr)-e^{2\varphi }\sqrt{1+e^{-2\varphi }\varphi\prsq}+\frac{\phi\prsq}{\sqrt{1+e^{-2\varphi }\varphi\prsq}} -e^{2\varphi } V( \phi )\right]\\
    &\quad \quad -\frac{1}{8\pi G_N}\int d\t dx\frac{d}{d\t}\left[e^\varphi\sinh^{-1}(e^{-\varphi}\varphi\pr)\right]
    \label{bdytrwq}
\end{align}

Finally, since $\abs{\dot{Z}}\simeq\abs{e^{-\varphi}\vp\pr}\ll 1$, we obtain the effective action of the EOW brane of type I as:
\begin{equation}
    I_{\rm{eff}} = \frac{1}{8\pi G_N}\int d\t dx\left(\frac{1}{2}(\partial_\t\varphi)^2 + (\partial_\t\phi)^2 - e^{2\varphi}(1 + V(\phi))\right).
\end{equation}
If the EOW brane ends on the asymptotic boundary at $t=0$ as in 
Fig.\ref{fig:hyperplane}, we canceled the boundary term (\ref{bdytrwq}) by adding the corner term (or Hayward term) \cite{Hayward:1993my,Boruch:2020wax,Boruch:2021hqs}. To deform the boundary condition we can further add the following boundary term
\begin{equation}
    I_{\rm{bdy}}=\frac{1}{8\pi G_N}\int dx \m_B e^{\vp},
\end{equation}
where $\mu_B$ is a constant which parametrizes the boundary condition 
as is familiar in the boundary Liouville theory \cite{Fateev:2000ik}. In this way, the total effective theory is identified as the Liouville field theory, which couples to another scalar field $\phi(\t)$ through its potential $V(\phi)$. Indeed, this action agrees with the Liouville CFT action with the central charge $c$ via the relation $c=\frac{3}{2G_N}$ \cite{Brown:1986nw}. 

We can process the same discussion for the type II case. Using $\e=\pm1$ defined in subsection \ref{subsec:generalArgInAdS3}, both type I and type II results are combined as
\begin{equation}\label{eq:EffLiouvilleAction}
    I_{\rm{eff}} = \frac{1}{8\pi G_N}\int_{\e \t<0} d\t dx\left[\frac{1}{2}\e(\partial_\t\varphi)^2 + (\partial_\t\phi)^2 - e^{2\vp}(\e+V(\phi))\right] + \frac{1}{8\pi G_N}\int_{\t=0} dx\m_B e^{\varphi}.
\end{equation}
In the type II case ($\ep=-1$) the overall sign of the action is opposite to that of the Liouville CFT. This is because it corresponds to the effective action when we integrate out a CFT with the central charge $c$ and this action is known to have the opposite sign \cite{Polyakov:1981rd,Caputa:2017urj,Caputa:2017yrh}.

We can easily construct a covariant expression of the action (\ref{eq:EffLiouvilleAction}) as follows:
\begin{equation}
  I_{\rm{cov}}\! =\! \frac{1}{8\pi G_N}\int d\tau dx \s{g}\left[
  -\frac{\ep}{8}R\frac{1}{\Box}R+(\de\phi)^2-(\ep+V(\phi))\right]
  \!+\! \frac{1}{8\pi G_N}\int_{\t=0} dx\m_B \s{\gamma}.    
\end{equation}

\subsection{Solving the equation of motion}
The equations of motion of  \eqref{eq:EffLiouvilleAction} are given as
\begin{align}
    \varphi \prpr &=-2\epsilon e^{2\varphi }( \epsilon +V( \phi )),\\
    2\phi \prpr &=-e^{2\varphi } \partial _{\phi } V( \phi )
\end{align}
These equations can be rewritten as
\begin{align}\label{eq:EffLiouvilleEOM1}
    V( \phi ) &=-\frac{1}{2} \epsilon e^{-2\varphi } \varphi \prpr -\epsilon ,\\
    \phi\prsq &= -\int e^{2\varphi }\frac{dV( \phi ( \tau ))}{d\tau } d\tau \label{eq:EffLiouvilleEOM2}
\end{align}
This means that, if we choose $\varphi(\t)$, the corresponding profile of the brane scalar $\phi(\t)$ and its potential $V(\phi)$ are uniquely determined. In addition, fields $\vp$ and $\phi$ are imposed boundary conditions:
\begin{equation}
    \partial_\t\phi = 0,\quad \partial_\t\vp + \m_B e^\vp = 0
\end{equation}
at $\t = 0$.

\subsection{Approximate Solution}
We start with the consistency check of the general argument in subsection \ref{subsec:generalArgInAdS3}. For simplicity, we consider the type II brane (i.e., $\e=-1$) defined by $z=Z(t)=\a t + \b t^p$ under $\a\ll 1$. In this case, since $\t$ is related to $t$ as $\t\simeq \sqrt{1-\a^2}t$, the Liouville field $\varphi(\t)$ is
\begin{equation}
    \varphi(\t) = -\log\left[\a\frac{\t}{\sqrt{1-\a^2}}+\b\left(\frac{\t}{\sqrt{1-\a^2}}\right)^p\right].
\end{equation}
Using the equations \eqref{eq:EffLiouvilleEOM1} and \eqref{eq:EffLiouvilleEOM2}, we obtain
\begin{equation}
    2(V(\phi)-1) \simeq \alpha ^{2} -\alpha \beta p( p-3)\left(\frac{\tau }{\sqrt{1-\alpha ^{2}}}\right)^{p-1},\quad \phi\prsq \simeq \frac{\beta p( p-1)}{2\alpha } \tau ^{p-3}.
\end{equation}
This leads the potential $V(\phi)$ as
\begin{equation}
    V(\phi) -1\simeq \frac{1}{2} \alpha ^{2} +\frac{( p-1)( 3-p)}{4} \alpha ^{2}\phi^{2},\label{eq:eff_mu_p}
\end{equation}
which agrees with \eqref{potoneg} to $\a^2$-order. The effective boundary cosmological constant $\m_B$ also turns out to be
\begin{equation}
    \m_B = \frac{\alpha }{\sqrt{1-\alpha ^{2}}} \simeq \a.
\end{equation}

In addition, using the conformal weight $\D$ \eqref{confdim(d=2)} of the boundary primary operator of BCFT inserted at $t=0$, the equation \eqref{eq:eff_mu_p} is reduced to
\begin{equation}
    V(\phi)-1 \simeq \frac{\ap^2}{2}+\D ( 1-\D )\phi^{2}.
\end{equation}
From this, we find that if the boundary primary is relevant ($\D<1$), the effective cosmological constant $V(\phi)$ increases during the time evolution with respect to $\t$. Conversely, if the boundary primary operator is irrelevant ($\D > 1$), $V(\phi)$ decreases. Note that both $V(\phi)-1$ and $\m_B$ at the boundary $\t=0$ are positive.

The EOW brane of type I can be discussed by the similar way. As the result, for the brane $z=-\a t-\b(-t)^p$, we obtain
\begin{align}
    \m_B &= -\a\\
    V(\phi) + 1 &= -\frac{\a^2}{2} + \D(1-\D)\phi^2,
\end{align}
which are negative at $t=\t=0$.

\subsection{Exact Solution}
There is an exact solution of a type II brane ($\e=-1$) to the equations \eqref{eq:EffLiouvilleEOM1} and \eqref{eq:EffLiouvilleEOM2}. This solution is obtained by setting
\begin{equation}
    \varphi(\t) = -\log(\b\t^p).
\end{equation}
Now $V(\phi)$ and $\phi(\t)$ are given as
\begin{equation}
    V(\phi) - 1=\frac{1}{2}p\b ^{2} \tau ^{2( p-1)},\quad \phi =\pm\sqrt{\frac{p( p-1)}{2}}\log \tau.
\end{equation}
This implies
\begin{equation}
    V(\phi) = 1+\frac{1}{2} p\b ^{2} e^{\pm\sqrt{2( 1-1/p)} \cdot 2\phi }.
\end{equation}
In addition, we find easily $\m_B$ vanishes. This potential corresponds to \eqref{eq:exponential_potential}.
%%%%%%%%%%%%%%%%%%%%%%%%%%%%%%%%%%%%%%%%%%%%%%%%%%%%%%%%
%%%%%%%%%%%%%%%%%%%%%%%%%%%%%%%%%%%%%%%%%%%%%%%%%%%%%%%%

%%%%%%%%%%%%%%%%%%%%%%%%%%%%%%%%%%%%%%%%%%%%%%%%%%%%%%%%
%%%%%%%%%%%%%%%%%%%%%%%%%%%%%%%%%%%%%%%%%%%%%%%%%%%%%%%%
\section{Cosmology from Branes}
\label{sec:cosmology}
%%%%%%%%%%%%%%%%%%%%%%%%%%%%%%%%%%%%%%%%%%%%%%%%%%%%%%%%
%%%%%%%%%%%%%%%%%%%%%%%%%%%%%%%%%%%%%%%%%%%%%%%%%%%%%%%%

The brane-world holography \cite{Randall:1999ee,Randall:1999vf,Karch:2000ct,Gubser:1999vj} argues that the bulk gravity surrounded by a EOW brane is dual to a lower dimensional gravity on the EOW brane with quantum effects taken into account, where the gravity gets localized. From this viewpoint, time-dependent type II brane solutions in AdS$_3$ can be regarded as a model for mini-cosmology. The classical brane dynamics in AdS$_3$ is expected to describe a two dimensional quantum gravity in cosmological spacetimes, such as the big bang and inflation. Therefore it is intriguing to explore cosmological solutions in our brane model. In order to have a global viewpoint of spacetime geometry, here we employ the global AdS$_3$ coordinate:
    \begin{align}
        ds^2=d\rho^2-\cosh^2\rho d\tau^2+\sinh^2\rho d\phi^2.  \label{gAdS}
    \end{align}
    We insert an EOW brane along the trajectory $\rho=R(\tau)$, which also extends in the $\phi$ direction.
    Two types of branes in the global AdS$_3$ are specified by
    \begin{align}
        \mathrm{Type \,I:}& \,\,\,R(\tau) \leq \rho, \notag \\
        \mathrm{Type \,II:}& \,\,\,0\leq\rho \leq R(\tau). \notag
    \end{align}
    
    We can write induced metric as follows:
    \begin{align}
        ds_Q^2=-(\cosh^2 R-\dot{R}^2)d\tau^2+\sinh^2 R d\phi^2, \label{gadsm}
    \end{align}
    where $\dot{R}=\frac{dR(\tau)}{d\tau}$. Since we assume that the EOW brane is time-like in the bulk AdS$_3$, we require $\cosh^2 R>\dot{R}^2$. The normal vector and extrinsic curvature are
    \begin{align}
        N=\frac{\epsilon\cosh R}{\s{\cosh^2R-\dot{R}^2}}(-d\rho+\dot{R}d\tau),
    \end{align}
    \begin{align}
        K_{ab}dx^adx^b
        =\epsilon\frac{\sinh R(\cosh^2R-2\dot{R}^2)+\ddot{R}\cosh R}{\s{\cosh^2R-\dot{R}^2}}d\tau^2
        -\epsilon\frac{\cosh^2R\sinh R}{\s{\cosh^2R-\dot{R}^2}}d\phi^2.
    \end{align}
    Again, $\epsilon=1$ and $\epsilon=-1$ corresponds to the type I and II, respectively.
    The boundary condition (\ref{KEOM}) can be written as follows:
    \begin{align}
     \dot{\phi}^2&=\epsilon\frac{(1-\sinh^2R)\dot{R}^2+\cosh R(\ddot{R}\sinh R-\cosh R)}{2\sinh R\s{\cosh^2R-\dot{R}^2}},\label{gKEOMP}\\
     V(\phi)&=-\epsilon\frac{\cosh^2R(\cosh^2R+\sinh^2R)-\dot{R}^2(2\sinh^2R+\cosh^2R)+\ddot{R}\sinh R\cosh R }{2\sinh R(\cosh^2R-\dot{R}^2)^{3/2}}.
    \label{gKEOMQ}    
    \end{align}
    The energy conservation like low is 
    \begin{align}
        \f{\sinh{R}}{\cosh^2{R}}\left( \f{\dot{\phi}^2}{\sqrt{\cosh^2{R}-\dot{R}^2}}+V(\phi) \sqrt{\cosh^2{R}-\dot{R}^2} \right) =-\epsilon
    \end{align}

\subsection{Brane cosmology}
    In order to construct a toy model of the expanding universe, we set a new coordinate by
    \begin{align}
        dT=\sqrt{\cosh^2R-\dot{R}^2}d\tau
    \end{align}
    which results in the FRW form induced metric
    \begin{align}
        ds_Q^2=-dT^2+a(T)^2d\phi^2
    \end{align}
    with the scale factor $a(T)=\sinh{R(\tau(T))}$.
    
    Let us derive an analogue of Friedmann equation.
    Here, we denote $a'(T)=\frac{da}{dT}$.
    The first and second derivative of $a$ are 
    \begin{align}
        a'(T)
        &=\cosh{R}\,\f{dR}{d\tau}\f{d\tau}{dT}
        =\frac{\dot{R}\cosh{R}}{\sqrt{\cosh^2R-\dot{R}^2}} \\
        a''(T)
        &=\f{d}{d\tau}\left( \frac{\dot{R}\cosh{R}}{\sqrt{\cosh^2R-\dot{R}^2}} \right) \f{d\tau}{dT}
        = \frac{\ddot{R}\cosh^3R-\dot{R}^4\sinh{R}}{{(\cosh^2{R}-\dot{R}^2)}^2}\label{gaddot}
    \end{align}
    To get the Friedmann equation we need the expression of the energy density $\varepsilon$ and pressure $p$ in terms of $R(\tau)$. Before that, however, we also need to recall that physical energy-momentum tensor is obtained by taking holographic renormalization. With this prescription, we subtract the value of $T_Q$ when the brane is exactly on the asymptotic boundary, here it is $\epsilon h_{ab}$ as we check soon. Thus the energy-momentum tensor is $T_{holo}^{ab}:=T_Q^{ab}-\epsilon h^{ab}$.
    Now, the energy density $\varepsilon$ and pressure $p$ are defined as 
    \begin{align}
        T_{holo}^{ab} \,( \,= \,T_Q^{ab}- \epsilon h^{ab}\,)
        =: \varepsilon\,e_\tau^a\, e_\tau^b+p\,e_\phi^a\, e_\phi^b \label{T_holo}
    \end{align}
     where vierbein $\{e_i^a\}$ satisfies $\eta_{ij}=h_{ab}e_i^a e_j^b$, equivalently $e_\tau^a \partial_a=\f{1}{\sqrt{\cosh^2{R}-\dot{R}^2}}\partial_\tau$, $e_\phi^a\partial_a = \f{1}{\sinh{R}}\partial_\phi$ and there is a completeness relation $h^{ab}=\eta^{ij}e_i^a e_j^b=-e_\tau^a e_\tau^b+e_\phi^a e_\phi^b$.
     Thus we get following expression
     \begin{align}
         \varepsilon-\ep \,
        % &=\,T^Q_{ab}e_\tau^a e_\tau^b 
        % = (K_{\tau\tau}-h_{\tau\tau}K)\left(\f{1}{\sqrt{\cosh^2{R}-\dot{R}^2}}\right)^2 
        &=(-\epsilon)\f{\cosh^2R}{\sinh{R}\sqrt{\cosh^2{R}-\dot{R}^2}}  \label{gepsiron} \\
        p+\ep \,
        % &=\, T^Q_{ab}e_\phi^a e_\phi^b 
        % = (K_{\phi\phi}-h_{\phi\phi}K)\left(\f{1}{\sinhR}\right)^2 
        &=(-\epsilon) \f{-\ddot{R}\cosh{R}+2\dot{R}^2\sinh{R}-\sinh{R}\cosh^2{R}}{(\cosh^2{R}-\dot{R}^2)^{3/2}}  \label{gp}
     \end{align}
     By combining \eqref{gaddot}, \eqref{gepsiron} and \eqref{gp},
     we see that the Friedmann equation becomes surprisingly simple form 
    \begin{align}
        \f{a''(T)}{a} \, \notag
        &=\,\f{\ddot{{R}}\cosh^3{R}-\sinh{R}(2\dot{{R}}^2\cosh^2{R}-\cosh^4{R}+(\cosh^2{R}-\dot{{R}}^2)^2)}{\sinh{R}(\cosh^2{R}-\dot{{R}}^2)^2} \\ \notag
        &=-1-\f{\cosh^2{R}}{\sinh{R}\sqrt{\cosh^2{R}-\dot{{R}}^2}} \f{-\ddot{{R}}\cosh{R}+\sinh{R}(2\dot{{R}}^2-\cosh^2{R})}{(\cosh^2{R}-\dot{{R}}^2)^{3/2}} \\ \notag
        &=-1-(-\epsilon)(\varepsilon-\ep)\,(-\epsilon)(p+\ep) \\ 
        &=-\ep(\varepsilon-p)-\varepsilon p. \label{Friedmann}
    \end{align}
    This is different from the well known Friedmann equation, it is because we used the Neumann boundary condition instead of the Einstein equation.
    
    As we noted in section\ref{sec:dSBranes}, pure AdS$_3$ satisfies $0=\nabla_a (K^{ab}-h^{ab} K)=\nabla_aT_Q^{ab}$. Inserting \eqref{T_holo} into the right hand side results in the equation of motion 
    \begin{align}
        \f{d\varepsilon}{d T}+\f{a'}{a}(\varepsilon +p)=0, \label{div-T}
    \end{align}
    which is the two dimensional version of ordinal cosmological equation of motion. 
    The above two equations \eqref{Friedmann} and \eqref{div-T} can be regarded as the fundamental equation of motion of this cosmology.
    Note that these equations do not depend on the bulk coordinate, in fact one can derive the same equation in \Poincare coordinate.

Below we would like to examine a few simple examples in the above cosmological interpretation. 

\subsection{dS brane solution}
First, consider the dS$_2$ brane which is a solution in the absence of scalar field. The profile is $R(\tau)=\mathrm{arcosh}\left[\f{\cosh{\xi}}{\cos{\tau}}\right]$. This is a solution of $\dot{\phi}^2=0,\,V=-\epsilon/\tanh{\xi}$, derived by using Hyperbolic-slice coordinate (see Appendix \ref{ap:hyp}). Energy density and pressure are
    \begin{align}
        \varepsilon=\ep\left(1-\f{1}{\tanh{\xi}}\right),\,\,\,p=\ep\left(-1+\f{1}{\tanh{\xi}}\right).
    \end{align}

\subsection{Constant Radius solution}

Next we would like to consider the constant radius solution $R(\tau)=b$, $\dot{\phi}^2$ has finite value so that the only possible setup is type II, and the boundary conditions \eqref{gKEOMP} \eqref{gKEOMQ} are written as follows;
\begin{align}
   \dot{\phi}^2 &= \frac{1}{2\tanh(b)}  \label{bphidot}\\ 
    V(\phi) &= \frac{1 }{ \tanh(2b)}\label{gtrivial}
\end{align}
The energy-momentum tensor is 
\begin{align}
    T^Q_{\tau\tau}&=\f{\cosh^2{b}}{\tanh{b}} \xrightarrow{b\to\infty} -h_{\tau\tau}  \\
    T^Q_{\phi\phi}&=-\sinh^2{b}\,\tanh{b} \xrightarrow{b\to\infty} -h_{\phi\phi} 
\end{align}
Under the $b\to\infty$ limit, the brane is exactly on the AdS boundary.
Here, the cosmological data are $a(T)=\sinh{b}$ and
\begin{align}
    \varepsilon=\f{1}{\tanh{b}}-1,&\,\,\,\,
    p=1-\tanh{b} .
\end{align}
They are positive and vanishes when the brane is on the AdS boundary, while $\dot{\phi}^2$ and $V$ survive.

\subsection{Big-bang like solutions}
It is also intriguing to consider solutions which describe a creation of universe via big-bang like phenomena. For this we focus on the small $R$ behavior of type II brane by setting 
$R(\tau)=0$ at $\tau=0$. We assume the profile 
\ba
R(\tau)\simeq A\tau^\ap,
\ea
in the limit $\tau\to 0$. We need to require $\ap\geq 1$ in order for the brane to be time-like. 

First let us examine the case $\ap>1$. Then we obtain from (\ref{gKEOMP}) 
\ba
\dot{\phi}^2\simeq V(\phi)\simeq \frac{1}{2A \tau^\ap}.
\ea

This leads to
\ba
\phi(\tau)\simeq \frac{2}{\s{2A}|2-\ap|}\tau^{-\frac{\ap}{2}+1}.
\ea
Then the potential looks like
\ba
V(\phi)\simeq \frac{1}{2A}\left(\s{2A}\frac{|2-\ap|}{2}\phi\right)^{\frac{2\ap}{\ap-2}}.
\ea
In this way, the potential gets positively divergent at $\tau=0$ i.e. the big-bang of our mini cosmology. Later the potential gets decreased and $R$ increases. In order to keep the potential finite at $\tau=0$, we need to consider the other possibility $\ap=1$ where we have $R(\tau)\simeq \tau$. The constant potential solutions, which will be studied in the next subsection, provide such an example.  

Now let us consider the other case $\alpha=1$. Adding the next order term, we assume the profile:
\ba
R(\tau)\simeq \tau -A \tau^p.
\ea
The time-like condition considered in the region $\tau \simeq 0$ is as follows:
\ba
A>0 &&(1<p<3) \\
1+6A>0&&(p=3)
\ea
When $p>3$, the EOW brane is automatically time-like near $\tau=0$. We obtain from (\ref{gKEOMP}) 
\ba
\dot{\phi}^2 \simeq
    \begin{cases}
    \frac{\sqrt{A p}(p+1)}{2\sqrt{2}}\tau^{\frac{p-3}{2}} &(1<p<3) \\
    \sqrt{1+6 A} & (p=3) \\
    1 & (3<p)
    \end{cases}
\ea
This leads to 
\ba
\phi \simeq
    \begin{cases}
    \left(\frac{4\sqrt{2A p}}{p+1}\right)^{\frac{1}{2}}\tau^{\frac{p+1}{4}} &(1<p<3) \\
    (1+6 A)^{\frac{1}{4}} \tau & (p=3) \\
    \tau & (3<p)
    \end{cases}
\ea
We also obtain from (\ref{gKEOMQ})
\ba
V(\phi) \simeq
    \begin{cases}
    \frac{3-p}{p+1}\phi^{-2} &(1<p<3) \\
    -\frac{3A^2-14A-2}{2(1+6 A)^{\frac{3}{2}}} - \frac{4212A^4+3024A^3-1116A^2-144A-1}{144(1+6A)^3}\phi^2 & (p=3) \\
    -\frac{A p(p-3)}{2}\phi^{p-5} & (3<p<5) \\
    -(5A-1)+\frac{10800A^2-480A+1}{144}\phi^2 & (p=5) \\
    1-\frac{Ap(p-3)}{2}\phi^{p-5} & (5<p<7) \\
    1-\frac{2016A-1}{144}\phi^2 & (p=7) \\
    1+\frac{1}{144}\phi^2 & (7<p)
    \end{cases}
\ea
If we add the higher term, $V(\phi)$ will change a little.

The cosmological scale factor is 
\begin{align}
    a(T)\simeq\tau\simeq\begin{cases}
        \left(\f{p-1}{\sqrt{4Ap}}T\right)^{\f{2}{p+1}}  & (1<p<3) \\
        \left(\f{2}{\sqrt{1+6A}}T\right)^{1/2} & (p=3) \\
        (2T)^{1/2} &(3<p)
    \end{cases}
\end{align}

%%%%%%%%%%%%%%%%%%%%%%%%%%
\subsection{Constant Potential Solutions}\label{sec:LorentzS}
%%%%%%%%%%%%%%%%%%%%%%%%%%

\begin{figure}[hhh]
    \centering
    \includegraphics[width=0.9\linewidth]{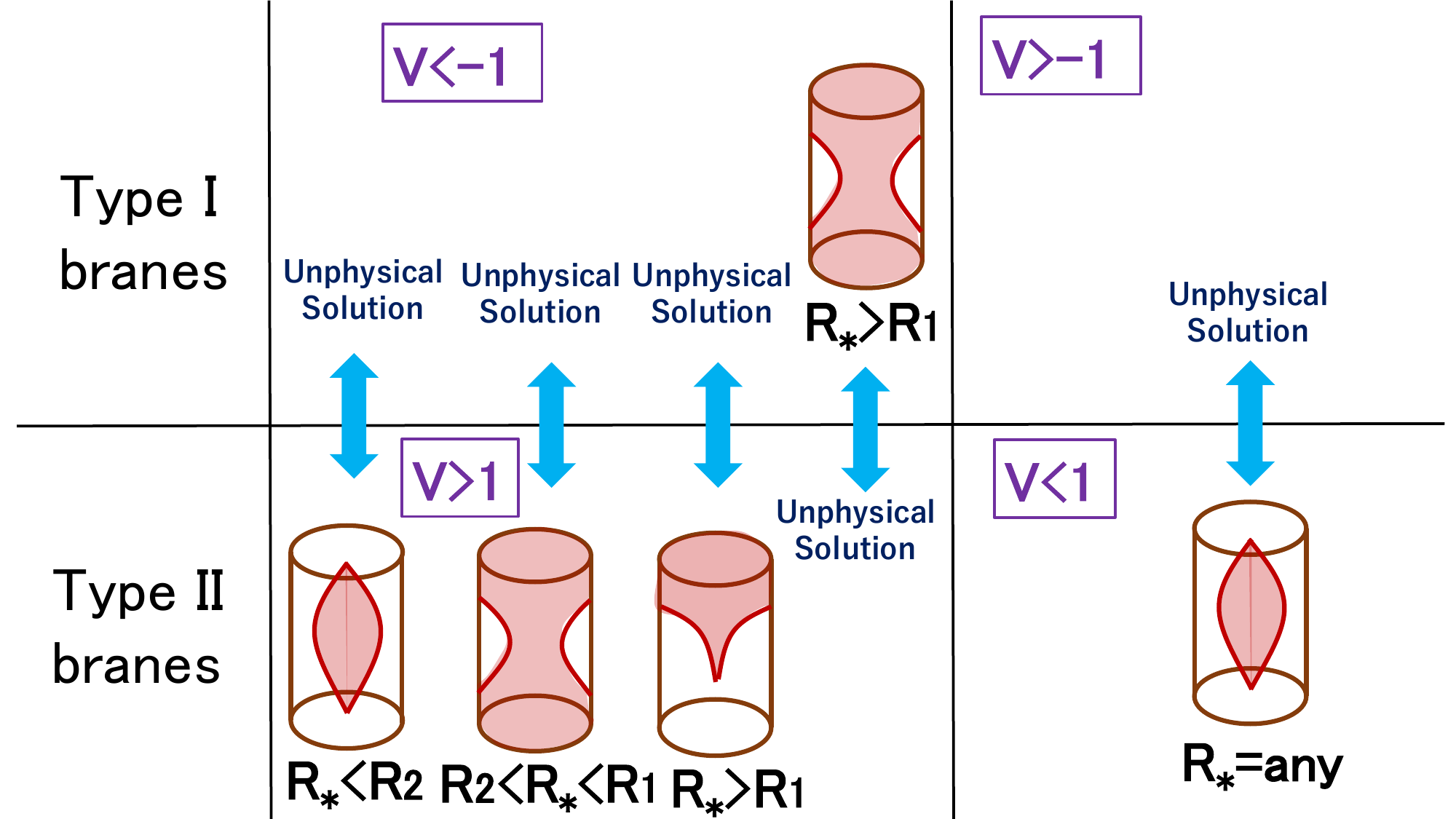}
    \caption{A sketch of profiles of physical solutions in type I (upper panels) and type II branes (lower panels). The red colored regions describe the physical spacetimes inside the EOW branes. The value $R_*$ denotes the radius where $\dot{R}$ vanishes.
    A physical solution in type I (or II) at $V=V_0$ corresponds to a unphysical solution (i.e. $\dot{\phi}^2<0$) at $V=-V_0$
  in type II (or I).  We set $R_2=\frac{R_1}{2}$. If $V>1$ and $R_*>R_1$ for type II case, there is no solution of $\dot{R}=0$. Instead we study $\dot{R}>0$ at $R=R_*$ in this regime.}
    \label{fig:phaselist}
\end{figure}

As an important model for the mini cosmology, we would like to focus on the case where the potential of the scalar field takes a constant value $V(\phi)=V$ and classify all possible solutions for both type I and type II branes. Below we will present our numerical analysis which exhausts all possible solutions. We sketched the summary of classification of solutions in Fig.\ref{fig:phaselist}. 
In fact, analytical solutions are also available. However, to make our presentation simpler, we will describe the analytical solutions in the appendix \ref{ap:Vzero} (when $V=0$) and appendix \ref{ap:Vnonzero} (when $V \neq 0$). We also present brane solutions in Euclidean AdS$_3$ in appendix \ref{sec:EuclidS}.

Before the explicit analysis, it is useful to first note that we need three initial conditions i.e. $R,\dot{R}$ and $\phi$ at a fixed time. However, the initial value of $\phi$ can be set to zero because of the obvious symmetry which shifts the value of $\phi$ as the potential is a constant. Moreover, as the system has the time translational invariance, we can always shifts a solution in the time direction. This removes one more parameter of the initial conditions. Thus non-trivial solutions are parameterized by only one parameter.

\subsubsection{Type II brane solutions}
First, we consider the type II brane solution in the global AdS$_3$ for a constant value of $V$ by numerically solving (\ref{gKEOMQ}). We require the scalar field is real valued to obtain physical solutions. An important class of solutions can be found by imposing the initial condition
$\dot{R}=0$ and $R=R_*$ at $\tau=0$. We find at $\tau=0$:
\begin{equation}
\ddot{R}=2 V\cosh^2 R_* -\frac{\cosh R_*(1+2\sinh^2 R_*)}{\sinh R_*}=\frac{\cosh R_*}{\sinh R_*}\left(V\sinh 2R_*-\cosh 2R_*\right),
\end{equation}
which leads to
\ba
\dot{\phi}^2=\frac{2\cosh^2 R_*(\cosh R_*-\sinh R_* V)}{2\sinh R_*}.\label{eq:constdphsq}
\ea
Since $\dot{\phi}^2\geq 0$, we get the important constraint at $\tau=0$:
\ba
V\leq \frac{\cosh R_*}{\sinh R_*}.  \label{constv}
\ea
Thus it is useful to introduce $R_1$ such that 
\ba
V=\frac{\cosh R_1}{\sinh R_1}.
\ea
Then for $R_*<R_1$, the condition (\ref{constv}) is satisfied.
We also find that for $R_*>\frac{R_1}{2}(\equiv R_2)$, we get $\ddot{R}>0$ and vice versa. 

\begin{figure}[htbp]
    \centering
    \includegraphics[width=0.3\linewidth]{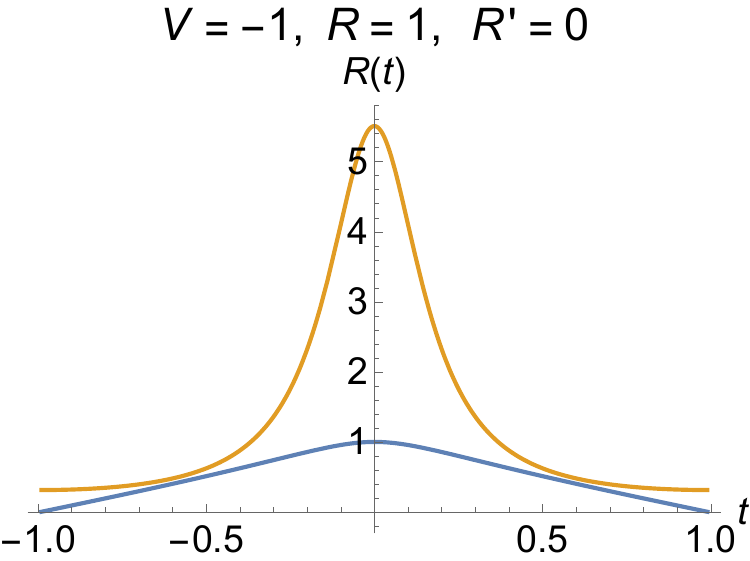}
 \includegraphics[width=0.3\linewidth]{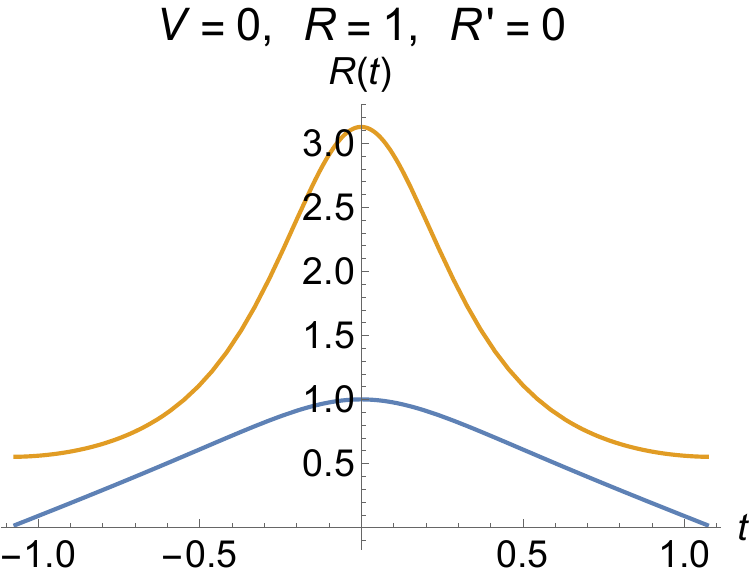}
     \includegraphics[width=0.3\linewidth]{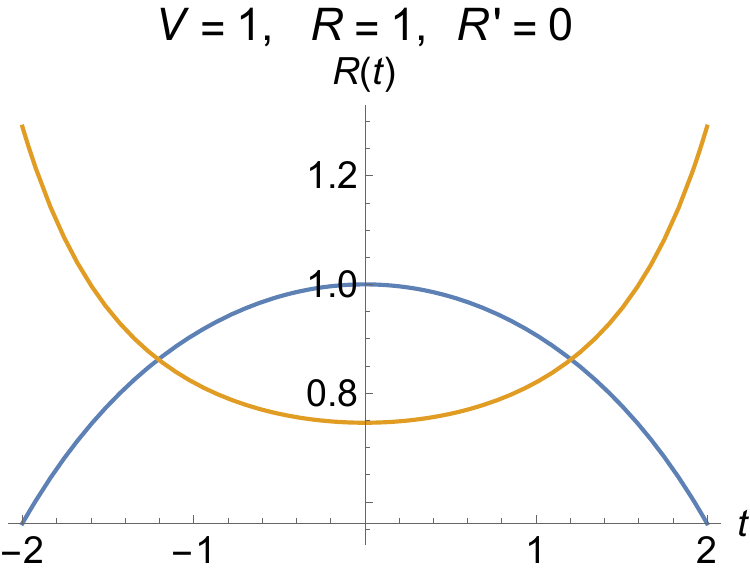}
    \caption{The plot of $R(\tau)$ (blue) and $\dot{\phi}^2$ (orange) as a function of $\tau$ for the initial condition $R=1$ and $\dot{R}=0$ at $\tau=0$.  We chose $V=-1$ (left), $V=0$ (middle) and $V=1$ (right). They are all bouncing universe solutions.}
    \label{fig:cosplotA}
\end{figure}

When $V<1$, for any value of $R_*$, the condition (\ref{constv}) is satisfied and always we have $\ddot{R}<0$ at $\tau=0$. Therefore we expect to have bouncing solutions, as depicted in Fig.\ref{fig:cosplotA}. Also this is consistent with the fact that in order to a type II brane to intersect at the AdS boundary we need $V\geq1 $, as we noted in section \ref{pertds}.

\begin{figure}[htbp]
    \centering
    \includegraphics[width=0.3\linewidth]{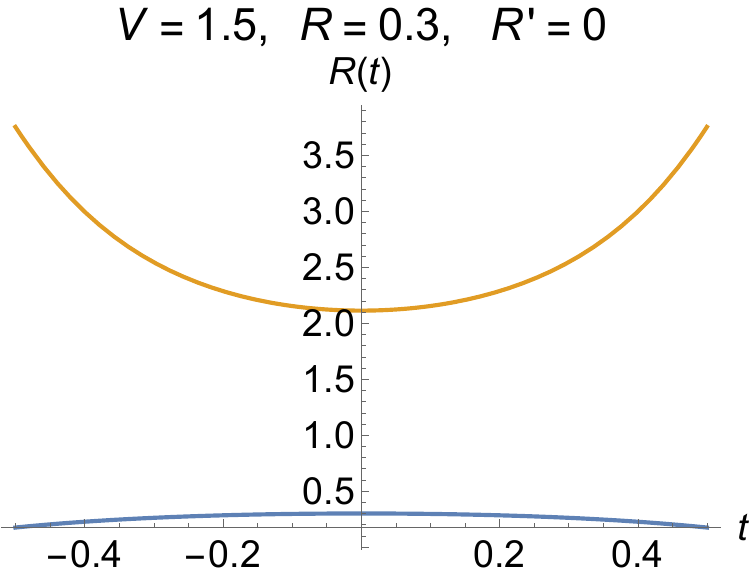}
   \includegraphics[width=0.3\linewidth]{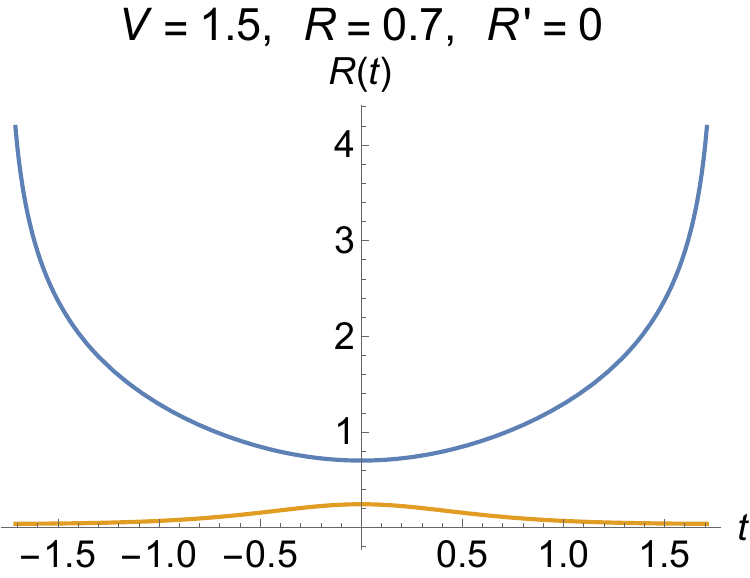}
 \includegraphics[width=0.3\linewidth]{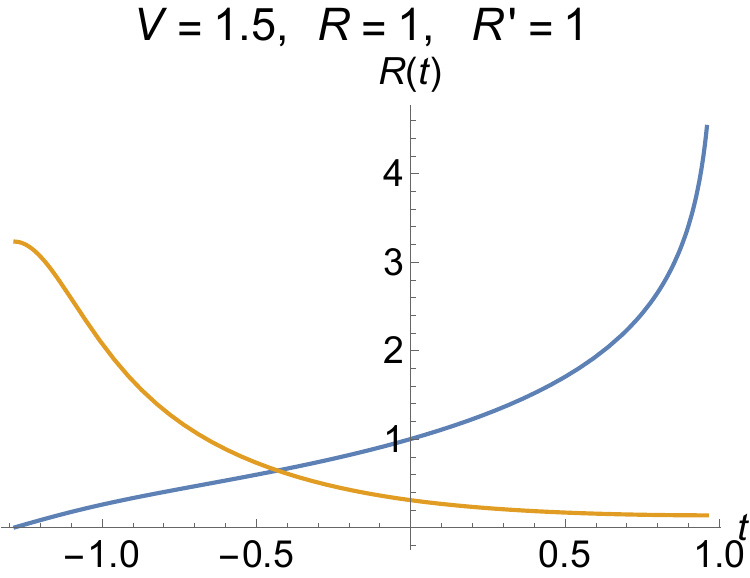}
    \caption{The plot of $R(\tau)$ (blue) and $\dot{\phi}^2$ (orange) as a function of $\tau$ for the constant potential $V=1.5$. We chose the initial conditions $(R,\dot{R})$ at $\tau=0$ of them to be $(0.3,0)$, $(0.7,0)$ and $(1,1)$ in the left, middle and right panel, respectively. Note that for $V=1.5$ we have  
    $R_1\simeq 0.805$  and $R_2\simeq 0.402$. The left graph describes a bouncing solution. In the middle one, the EOW brane reaches the AdS boundary in the past and future. The right one corresponds to a big-bang like solution where universe is created in the past.}
    \label{fig:cosplotB}
\end{figure}

When $V\geq 1$, the brane can reach the AdS boundary $R=\infty$ at a finite time, as in the dS$_2$ branes. We find that for $R_*<\frac{R_1}{2}$ we get the bouncing solution as shown in the left panel of Fig.\ref{fig:cosplotB}.
, while for $\frac{R_1}{2}<R_*<R_2$ we get a time symmetric solution which reaches the AdS$_3$ boundary at both past and future (finite) time as in the middle panel of Fig.\ref{fig:cosplotB}. For $R_*>R_1$, we cannot find any solution with $\dot{R}=0$ at a time. Instead we can find a solution which starts from a big bang like creation of universe i.e. $R=0$ at an initial time and then expands toward the AdS boundary within a finite time, depicted in the right panel of Fig.\ref{fig:cosplotB}.

To better understand the properties of big-bang like solutions, it is helpful to examine the behavior near $R=0$. By studying the solutions near $R=0$, we always find the behavior $R(\tau)\simeq \tau$. By solving (\ref{gKEOMQ}) as a power series of $\tau$ we obtain
\ba
&& R(\tau)=\tau-A\tau^3+\frac{1}{10}(-2-14A+3A^2+2+2\s{1+6A}+12A\s{1+6A}V)+O(\tau^7),\no
&&\dot{\phi}^2=\s{1+6A}-\frac{2}{\s{1+6A}}\left(-1-6A+\s{1+6A}V+6A\s{1+6A}V\right)+O(\tau^4).  \label{bigb}
\ea
It is clear that this is a physical solution $\dot{\phi}^2>0$ for any $A> -\frac{1}{6}$.  $A=-\frac{1}{6}$ corresponds to the null solution where the brane becomes light-like.

\subsubsection{Type I brane solutions}

Now we turn to the constant potential solution of type I branes. We can relate type I brane solutions to the previous analysis of type II by replacing $V$ with $-V$ and $\dot{\phi}^2$ with $-\dot{\phi}^2$. The latter means that a real value scalar field in type I (or II) becomes an imaginary one in type II (or I). Below we focus on the real valued solutions for type I branes.

Again let us try to find solutions under the initial condition $\dot{R}=0$ and $R=R_*$ at $\tau=0$. Then we obtain
\ba
\ddot{R}&=&-\frac{\cosh R_*}{\sinh R_*}\left(V\sinh 2R_*+\cosh 2R_*\right),\no
\dot{\phi}^2&=&-\frac{2\cosh^2 R_*(\cosh R_*+\sinh R_* V))}{2\sinh R_*}.
\ea
The non-negativity of the above equation gives the important constraint
\ba
V\leq -\frac{\cosh R_*}{\sinh R_*}.  \label{constav}
\ea
Thus it is useful to introduce $R_1$ such that 
\ba
-V=\frac{\cosh R_1}{\sinh R_1}.
\ea
Then we find $\ddot{R}>0$  if $R_*>\frac{R_1}{2}$ and vice versa. 

\begin{figure}[H]
    \centering
    \includegraphics[width=0.45\linewidth]{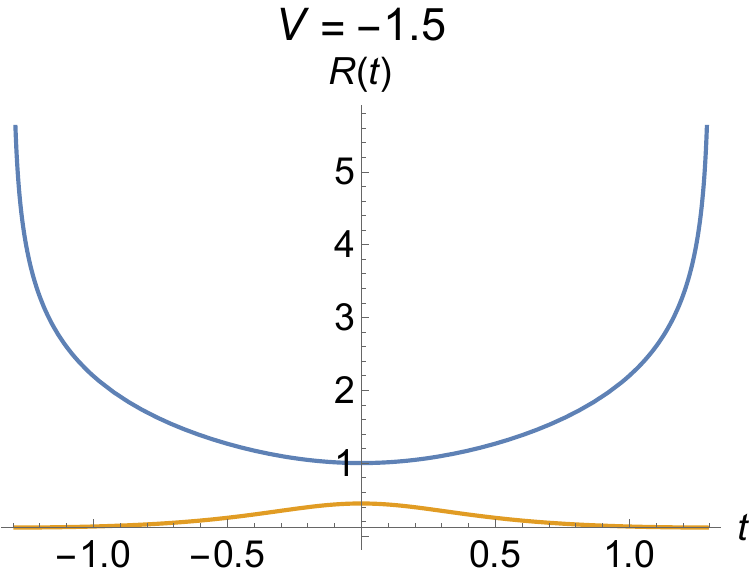}
    \caption{The plot of $R(\tau)$ (blue) and $\dot{\phi}^2$ (orange) as a function of $\tau$ for the type I brane with the initial condition $R=1$ and $\dot{R}=0$ at $\tau=0$ for $V=-1.5$.}
    \label{fig:cosplotQQW}
\end{figure}

In order to find a real valued solution we need $V\leq -1$ as is clear from (\ref{constav}). We can numerically confirm that when $V>-1$, there is no solution at all. Thus we focus on the $V\leq -1$ case. For $R_*>R_1$, we can find a time symmetric solution where we have $R=\infty$ at $\tau=\tilde{\tau}$ and $\tau=-\tilde{\tau}$, where $\tilde{\tau}$ is finite. This is depicted in Fig.\ref{fig:cosplotQQW}. We cannot find any other types of solutions in type I brane for a constant $V$. Indeed, there is no big-bang like solution of the form (\ref{bigb}) as the sign of $\dot{\phi}^2$ flips.

We summarized the profiles of type I and II branes in Fig.\ref{fig:phaselist}. It is useful to note that in order to end on the AdS$_3$ boundary we need $V\leq -1$ for type I brane and $V\geq 1$ for type II brane. At $V=-1$, there is only the constant scalar solution in type I, while at $V=1$, there is non-trivial solution with scalar in type II.

%%%%%%%%%%%%%%%%%%%%%%%%%%%%%%%%%%%%%%%%%%%%%%%%%%%%%%%%
%%%%%%%%%%%%%%%%%%%%%%%%%%%%%%%%%%%%%%%%%%%%%%%%%%%%%%%%
\section{EOW Branes with boost symmetry}
\label{sec:boost}
%%%%%%%%%%%%%%%%%%%%%%%%%%%%%%%%%%%%%%%%%%%%%%%%%%%%%%%%
%%%%%%%%%%%%%%%%%%%%%%%%%%%%%%%%%%%%%%%%%%%%%%%%%%%%%%%%
    
    In this section, we assign boost symmetry on EOW branes in Poincar\'{e} coordinate, while we have assigned translational symmetry in section \ref{sec:dSBranes}.
    The EOW brane is defined as hypersurface $z=Z(\rho)$, where $\rho=\sqrt{|t^2-x^2|}$. To distinguish whether the brane lies in region $t^2>x^2$ or $x^2>t^2$, we introduce $\sigma:=\mathrm{sign}(t^2-x^2)$. 
    In each region, the metric and induced metric are 
    \begin{align}
        ds^2=\f{dz^2-\sigma(d\rho^2-\rho^2d\eta^2)}{z^2}
        \to \f{-\sigma(1-\sigma Z'^2)d\rho^2+\sigma\rho^2d\eta^2}{Z^2}
    \end{align}Here we denoted $Z':=\f{dZ}{d\rho}$.
    Note that when $\sigma=1$ the time-like condition requires $1-Z'^2>0$, but when $\sigma=-1$ the brane is automatically time-like and there is no extra condition on the function $Z(\rho)$.
    
    Normal vector is 
    \begin{align}
        N=\chi\f{dz-Z'd\rho}{Z\sqrt{1-\sigma Z'^2}}.
    \end{align}
    Here we express the direction of normal vector using $\chi=\pm1$. This works in the same way as $\epsilon=\pm1$ have done so far, but its physical meaning is ambiguous since we cannot classify this setup based on whether it has AdS boundary or not.
    The normal vector leads to the extrinsic curvature
    \begin{align}
        K_{ab}dx^adx^b=\chi\f{\{\sigma(1-\sigma Z'^2)-ZZ''\}d\rho^2+(-\sigma\rho^2+\rho ZZ'')d\eta^2}{Z^2\sqrt{1-\sigma Z'^2}}.
    \end{align}
    Therefore the Neumann boundary condition is
    \begin{align}
        {\phi'}^2
        &=\f{1}{2}(K_{\rho\rho}-h_{\rho\rho}h^{\eta\eta}K_{\eta\eta})
        =\chi\f{(1-\sigma Z'^2)Z'-\rho Z''}{\rho Z\sqrt{1-\sigma Z'^2}}\\
        V&=\f{1}{2}K
        =\f{\chi}{2\rho(1-\sigma Z'^2)^{3/2}}[\sigma\rho ZZ''+(\sigma ZZ'-2\rho)(1-\sigma Z'^2)].
    \end{align}

    The NEC is satisfied when $(1-\sigma Z'^2)Z'-\rho Z''=0$, which is easily solved as
    \begin{align}
        Z'(\rho)=\f{\rho A}{\sqrt{1+\sigma\rho^2 A^2}}
    \end{align}
    where $A$ is constant. As an generalization we allow $A$ to be function $A=A(\rho)$, which leads to boundary condition
    \begin{align}
        {\phi}'^2&=\f{-\chi\rho A'(\rho)}{2Z(\rho)(1+\sigma\rho^2 A^2(\rho)) },\label{phid2_A}
    \end{align}
    and some more complicated expression of $V$.

    In this setup, let us discuss some examples which can be constructed by analytical continuation of Euclidean versions \cite{Kanda:2023zse}.

    \noindent\underline{Floating `sphere'}
    
    When we define the brane as $Q: -t^2+x^2+(Z-a)^2=R^2$, the boundary condition is solved as 
    \begin{align}
        {\phi}'^2=0,\,\,V=\f{a}{R},
    \end{align}
    so this is a consistent solution.
    This brane contains the two region $\sigma=\pm1$, in both region arbitral function is related to the radius $A=\f{1}{R}$.

    \noindent\underline{`cone'}

    Defining as $Q: Z(\rho)=a\rho$, we think of it as reside in region $\sigma=-1$. 
    The normal vector and boundary conditions are
    \begin{align}
        N&=\chi\f{dz-ad\rho}{a\rho\sqrt{1+ a^2}}, \\
        {\phi}'^2
        &=\chi\f{\sqrt{1+a^2}}{\rho^2},\,\,
        V=-\chi\f{(2+a^2)}{2(1+a^2)^{1/2}}.
    \end{align}
    Thus $\chi=+1$, which concludes that the spacetime resides between the brane and AdS boundary.
    Note that the singularity at $\rho=0$ is contained in the arbitral function $A=\f{a}{\rho\sqrt{1+a^2}}$

    \noindent\underline{Floating `torus'}

    Another example is $Q: (\rho-R)^2+(Z(\rho)-b)^2=a^2 \Leftrightarrow Z(\rho)=b\pm\sqrt{a^2-(\rho-R)^2}$.
    We choose $\sigma=-1$.
    The normal vector is 
    \begin{align}
        N=\f{\chi}{Z\sqrt{1+Z'^2}}\left(dz+\f{\rho-R}{Z-b}d\rho\right)
    \end{align}
    The arbitral function is $A(\rho)=\pm\f{R-\rho}{a\rho}$, where $\pm$ coincides with the sign of $Z(\rho)-b$.
    One boundary condition \eqref{phid2_A} is 
    \begin{align}
        {\phi}'^2= \pm\chi\f{aR}{2\rho Z(Z-b)^2}.
    \end{align}
    So $\chi=+1 $ for $Z-b>0$ and $\chi=-1$ for $Z-b<0$, which indicates that the spacetime is inside of the `torus'.
    Potential is 
    \begin{align}
        V(\rho)=\f{b}{a}-\f{R}{2a\rho}(b\pm\sqrt{a^2-(\rho-R)^2})
    \end{align}

    \begin{figure}[H]
      \begin{minipage}[b]{0.28\columnwidth}
        \centering
        \includegraphics[width=\columnwidth]{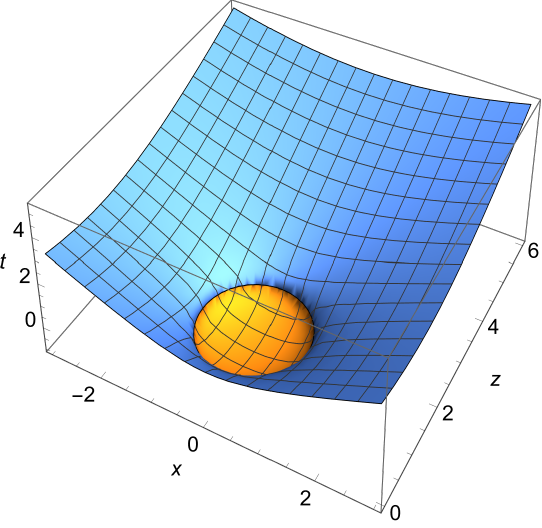}
        % \caption{sphere}
        \label{fig;sphere}
      \end{minipage}
      \hspace{0.03\columnwidth} % ここで隙間作成
      \begin{minipage}[b]{0.28\columnwidth}
        \centering
        \includegraphics[ width=\columnwidth]{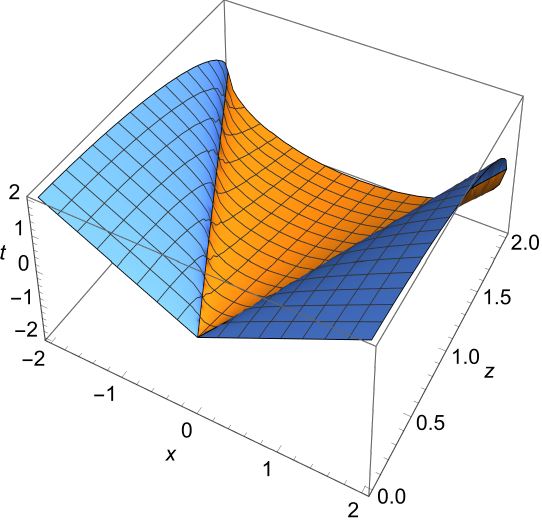}
        % \caption{cone}
        \label{fig;cone}
      \end{minipage}
      \hspace{0.03\columnwidth} % ここで隙間作成
      \begin{minipage}[b]{0.28\columnwidth}
        \centering
        \includegraphics[ width=\columnwidth]{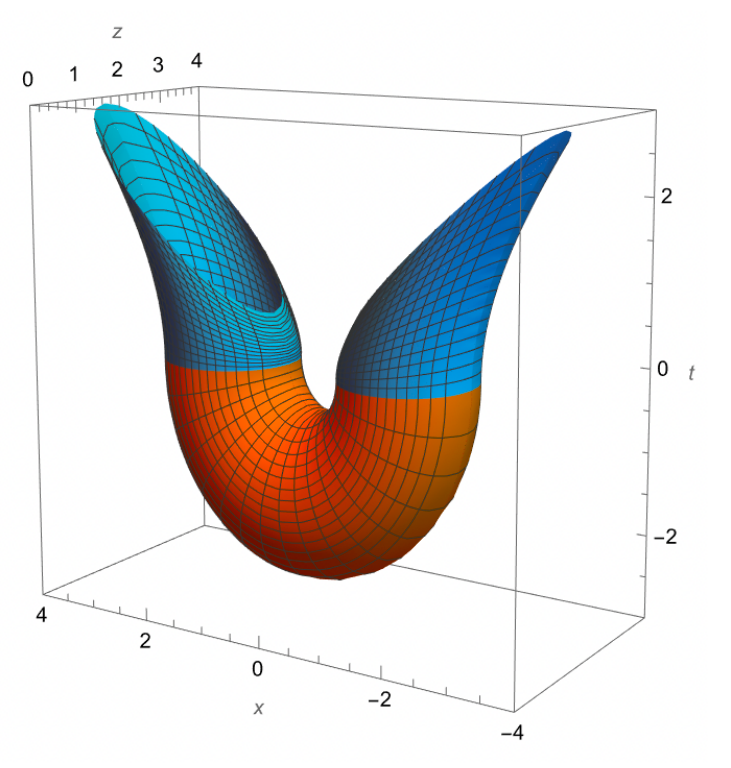}
        % \caption{torus}
        \label{fig;torus}
      \end{minipage}
      \caption{Plots of analytically continuated branes. The blue part ($t>0$) are our solutions and the orange part ($t<0)$ are corresponding Euclidean branes.
      (Left) The `sphere' is described by setting $(a,R)=(1.5,1.0)$, we can freely choose the spacetime to be upper or lower side of the brane.  
      (Center) The `cone' is described by setting $a=1.0$, its spacetime is lower side of the brane.
      (Right) The `torus' is described by setting $(a,b,R)=(1.0,1.5,1.5)$, its spacetime is in the `torus'.}
    \end{figure}

%%%%%%%%%%%%%%%%%%%%%%%%%%%%%%%%%%%%%%%%%%%%%%%%%%%%%%%%
%%%%%%%%%%%%%%%%%%%%%%%%%%%%%%%%%%%%%%%%%%%%%%%%%%%%%%%%
\section{Conclusions and Discussions}
\label{sec:conclusions}
%%%%%%%%%%%%%%%%%%%%%%%%%%%%%%%%%%%%%%%%%%%%%%%%%%%%%%%%
%%%%%%%%%%%%%%%%%%%%%%%%%%%%%%%%%%%%%%%%%%%%%%%%%%%%%%%%

In this paper, we studied the time-dependent dynamics of end-of-the-world (EOW) branes in AdS backgrounds, mainly focusing on the three dimensional AdS (AdS$_3$). Our important ingredient is to put a scalar field $\phi$ on the EOW brane, though the bulk gravity is the pure gravity, which allows us to keep the theory solvable. 
We note that the bulk gravity couples to the localized scalar field $\phi$ through the Neumann boundary condition imposed on the brane. Thus, we can see the scalar field changes brane setups, which enables us to extract rich brane dynamics, which looks like a mini-cosmology.

When the brane intersects with the AdS boundary, this provides an example of the AdS/BCFT correspondence. When it does not, we can view this as a quantum gravity model of mini-cosmology on the brane, dual to the classical gravity in the bulk region surrounded by the brane, assuming brane-world holography. In the former case, there are two types of brane configurations: type I and type II as depicted in Fig.\ref{fig:setup}, while in the latter case, we can only allow the type II brane.

When the brane intersects with the AdS boundary, the bulk gravity surrounded by the brane and AdS boundary is dual to a boundary conformal field theory (BCFT) on a lower half plane as shown in Fig.\ref{fig:hyperplane}, which has a boundary at $t=0$. Such a setup with a space-like boundary has not been studied well till now because it violate basic properties of quantum field theories such as the causality. However, a space-like boundary in the type I brane can be regarded as a final state projection and is expected to be important in order to embed quantum information theoretic operations into quantum field theories. In the type II brane, we may more properly interpret the setup as a CFT on a lower half plane coupled to a gravity in a dS. We believe these themselves deserve future studies also from field theory sides, for which our holographic results provide predictions. 

Without the localized scalar field, the time-like world-volume of the brane becomes a de Sitter space (dS), assuming the translational invariance in the space direction. Thus we can regard the scalar field as a matter field in the dS gravity. We used this setup to embed the dS/CFT into a higher dimensional AdS/CFT. We found that if we assume the null energy condition in this embedding in AdS/CFT, then we have a strong constraint on the spectrum in dS/CFT such that the conformal dimension takes real values and the mass $M$ of a scalar field on the dS$_d$ with a radius $R_{dS}$ is bounded as $M^2R^2_{dS}\leq \frac{(d-1)^2}{4}$. We also construct fully back-reacted solutions, with various scalar field profiles, each of which can be viewed as examples of the dS/CFT.
In order to see the relation between the EOW brane setups and the potentials, we put figures of them in each example.

In the BCFT side, the scalar field perturbations for type I branes can be regarded as RG flows. We derived a monotonicity property in field theories with a space-like boundary, which is analogous to the $g$-theorem in the time-like boundary case, by employing our holographic descriptions with the null energy condition imposed. Interestingly, this time-like $g$-theorem works also for type II brane, which predicts a monotonicity property in a system of a CFT coupled to a dS gravity. 

In the latter half of this paper, we investigated cosmological properties of our brane model in AdS$_3$. First we noted that when the curvature of the brane is small, the effective action of the two dimensional quantum gravity on the brane is given by the Liouville gravity with a scalar field matter. This clarifies how the brane-world holography looks like in our brane model.

In order to make mini-cosmology models, we extensively studied the time-dependence of brane solutions in global AdS$_3$ geometry. We were able to rewrite the equation of motion of branes in terms of Friedman like equation, which governs how the two dimensional spacetime evolves, by introducing energy density and pressure. We also found that the big-bang like solutions are possible for suitable choices of the potential $V(\phi)$. We also exhaust all solutions when the potential $V$ takes a constant value as summarized in Fig.\ref{fig:phaselist}, by both numerical and analytical calculations. For type II branes, we can have varieties of solutions depending on the value of $V$: (a) a universe is created by a big-bang and is later shrunk to zero size, ended up with a big-crunch, (b) a time symmetric universe which has a minimum radius in the middle, and (c) after a universe is created at the big-bang, it expands to a infinite size one to reach the AdS boundary. For type I branes, we can only find a universe which is created at the AdS boundary, expands for a while and finally disappears by intersecting again with the AdS boundary.  

It will be an interesting future problem to push such brane models further in higher dimensions towards constructing realistic models of cosmology. In particular, it would be intriguing to better understand the quantum gravity effect in big-bang singularities as our holographic model can describe a quantum gravity in spacetime with a singularity in terms of a classical gravity in a smooth higher dimensional spacetime with a conical brane. It is intriguing to note that the conical singularities in BTZ black holes turn out to be deformed to a geometry with a horizon in the AdS$_4$ brane-world model \cite{Emparan:2020znc}. This may motivate us to study our brane with the scalar field in higher dimensions.

%%%%%%%%%%%%%%%%%%%%%%%%%%%%%%%%%%%%%%%%%%%%%%%%%%%%%%%%
%%%%%%%%%%%%%%%%%%%%%%%%%%%%%%%%%%%%%%%%%%%%%%%%%%%%%%%%
\section*{Acknowledgements}

%%%%%%%%%%%%%%%%%%%%%%%%%%%%%%%%%%%%%%%%%%%%%%%%%%%%%%%%
%%%%%%%%%%%%%%%%%%%%%%%%%%%%%%%%%%%%%%%%%%%%%%%%%%%%%%%%

This work is supported by MEXT KAKENHI Grant-in-Aid for Transformative Research Areas (A) through the ``Extreme Universe'' collaboration: Grant Number 21H05187. TT is also supported by Inamori Research Institute for Science, and by JSPS Grant-in-Aid for Scientific Research (A) No.~21H04469. HK is supported by JSPS KAKENHI Grant Number 24KJ1476.

\appendix

\section{Analytical Solutions at $V(\phi)=0$}\label{ap:Vzero}
Let us consider the case $V(\phi)=0$ in the type II brane in global AdS$_3$. By setting $V=0$ in (\ref{gKEOMQ}), we find
\ba
\ddot{R}=\frac{2\sinh^2 R+\cosh^2 R}{\sinh R \cosh R}\dot{R}^2-\frac{\cosh R}{\sinh R}(\cosh^2 R+\sinh^2 R),
\ea
where $\dot{Q}=\frac{dQ}{dt}$.
We can rewrite this as 
\ba
\frac{d}{dt}\left(\frac{\dot{R}}{\cosh^2 R \sinh R}\right)=-\frac{\cosh^2 R+\sinh^2 R}{\sinh^2 R \cosh R}.
\label{xxxq}
\ea

By introducing $y(t)$ such that 
\ba
\dot {y}=\frac{\dot{R}}{\cosh^2 R \sinh R}, \label{yuuhe}
\ea
the equation of motion (\ref{xxxq}) can be rewritten as 
\ba
\ddot{y}=-\frac{\cosh^2 R+\sinh^2 R}{\sinh^2 R \cosh R}.  \label{eomghh}
\ea
We can solve (\ref{yuuhe}) as 
\ba
y=\log \tanh\frac{R}{2}+\frac{1}{\cosh R}.
\ea
In $R\to 0$, we find
$y\simeq 1+\log\frac{R}{2}\to -\infty$.
In $R\to \infty$ limit we have $y\simeq -\frac{8}{3}e^{-3R}$.

Then (\ref{eomghh}) can be rewritten as 
\ba
\ddot y=-\frac{dU(y)}{dy},
\ea
where the potential $U(y)$ is given by 
\ba
U(y)=-\frac{1}{2\sinh^2 R \cosh^2 R}.
\ea
In the $y\to 0$ limit we find 
$U(y)\simeq -8\left(-\frac{3y}{8}\right)^{\frac{4}{3}}$.
On the other hand, in the limit $y\to \infty$ we obtain
$U(y)\simeq -\frac{1}{8}e^{2-2y}$.

\begin{figure}[htbp]
    \centering
    \includegraphics[width=0.6\linewidth]{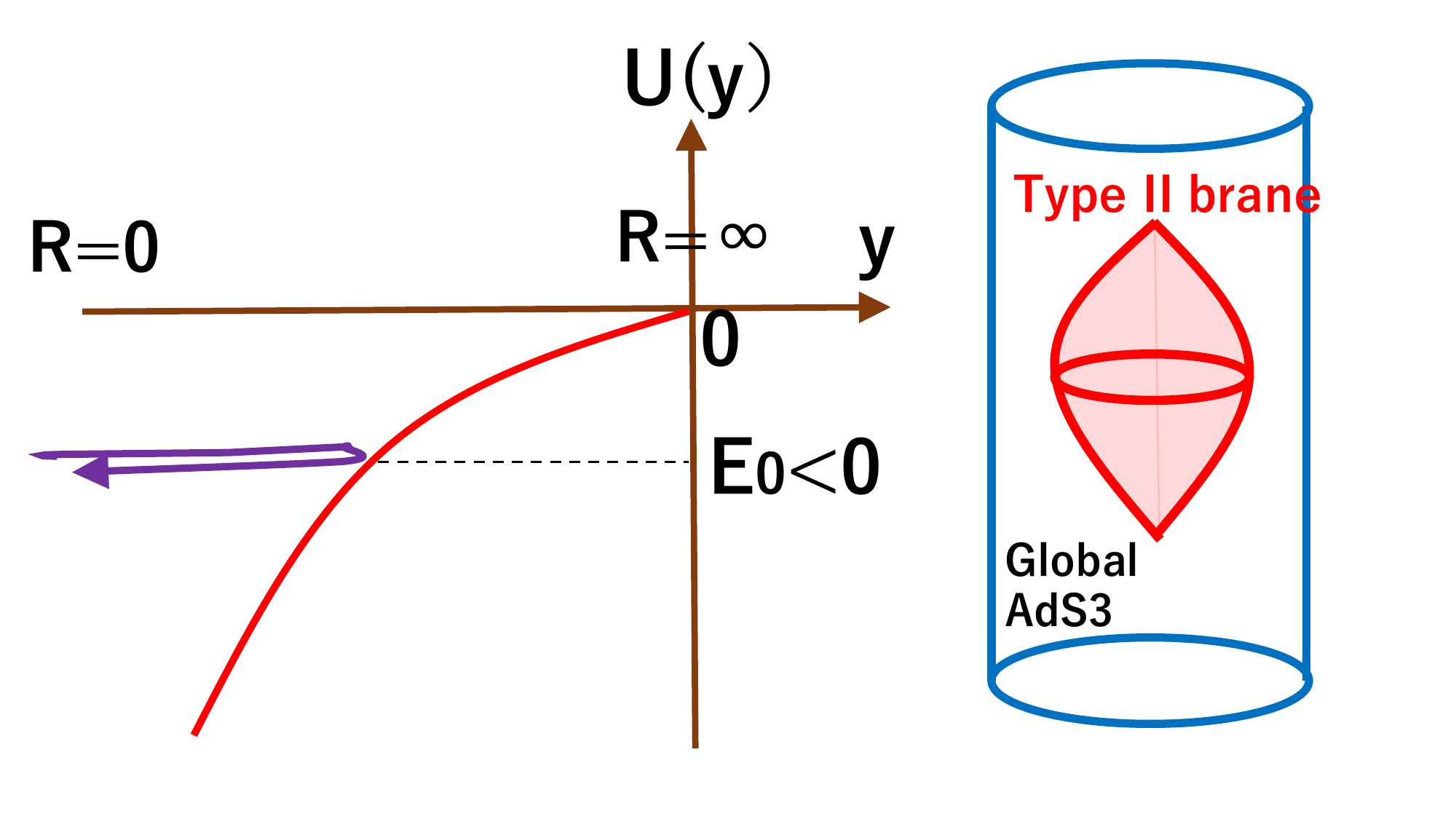}
    \caption{The sketch of the bounce solution in the potential $U(y)$ (left panel) and the type II brane in  global AdS$_3$ (right panel). }
    \label{fig:potential}
\end{figure}

The energy conservation tells us 
\ba
\frac{1}{2}\dot{y}^2+U(y)=E_0. 
\ea
For this trajectory we can evaluate 
\ba
\cosh^2 R-\dot{R}^2=-2E_0\cosh^4 R \sinh^2 R.  \label{confgq}
\ea
Since this should be positive to get the time-like brane and this requires $E_0<0$. Thus we get the bounce solution as sketched in the left panel of Fig.\ref{fig:potential}. This describes a closed universe with big-bang and big-crunch singularity as depicted in the right panel of Fig.\ref{fig:potential}.

The scalar field profile is also given by 
\ba
\dot{\phi}^2=\frac{\cosh R^2}{\sinh R}\s{\cosh R^2-\dot{R}^2}=\s{-2E_0}\cosh^4 R.
\ea

Finally we turn to the case $V((\phi)=0$ in the type I case. The equation of motion of $R(t)$ is identical to the type II one and we get (\ref{confgq}). However, for the type I brane, we find $\dot{\phi}^2<0$. Thus there is no real value solution in the type I case.

\section{Analytical Solutions at $V(\phi)\neq 0$}\label{ap:Vnonzero}

In this section, we will explore the analytical solution for $V\neq 0$ of type II brane. To do this, it is convenient to parametrize the brane as $t=T(\rho)$, since the equation of motion of the EOW brane, a second-order differential equation with respect to $T(\rho)$, does not explicitly depend on $T(\rho)$ itself due to time-translational symmetry of global AdS. In other words, the equation of motion is just a first-order differential equation with respect to $S(\rho):=T\pr(\rho)$.

The induced metric on the brane is given by
\begin{equation}
    ds^2 = -(S^2\cosh^2\r-1)d\r^2 + \sinh^2\r d\phi^2.
\end{equation}
Because the EOW brane is a time-like surface, we demand $S\cosh\r\geq1$. Note that, due to time-reversal symmetry, it is safe to assume $S>0$ without loss of generality. For later convenience, we define
\begin{equation}
    \cosh H(\r):=S(\r)\cosh\r.
\end{equation}
Using $H(\r)$ the normal vector is taken as
\begin{equation}
    N^\m = \left(\frac{1}{\tanh H},\frac{1}{\sinh H\cosh\r},0\right).
\end{equation}
Additionally, the extrinsic curvature and the stress tensors are
\begin{equation}
    K_{\r\r} - Kh_{\r\r} = \frac{\sinh (2H) }{2\tanh \rho },\quad K_{\phi\phi} - Kh_{\phi\phi} = \frac{\sinh^{2} \rho }{\sinh^{2} H }\left(H\pr -\frac{1}{2}\tanh \rho \sinh (2H) \right)
\end{equation}
and
\begin{equation}
    T_{\r\r} = \phi\prsq +V\sinh^{2} H,\quad T_{\phi\phi} = \frac{\sinh^{2} \rho }{\sinh^{2} H }\left(\phi\prsq -V\sinh^{2} H \right).
\end{equation}
Hence, we obtain the equation of motion of the brane
\begin{equation}
    H\pr = \frac{\sinh (2H) }{\tanh (2\rho) } -2V\sinh^{2}H.
\end{equation}
Fortunately, this equation reduces to
\begin{equation}
    \frac{d}{d\r}\left(V\cosh( 2\rho )-\frac{\sinh( 2\rho )}{\tanh H}\right)=0.
\end{equation}
Therefore, using some constant $V_0$, which will be determined later, we find the solution
\begin{equation}\label{eq:H_sol_const_V}
    V \cosh(2\r)-\frac{\sinh(2\r)}{\tanh H} = V_0.
\end{equation}
Finally, $T(\r)$ is obtained by integrating $S(\r)$:
\begin{equation}
    T(\r) = \int d\r S(\r) = \int d\r\frac{V\cosh( 2\rho ) -V_0}{\cosh \rho \sqrt{( V\cosh( 2\rho ) -V_0)^{2} -\sinh^{2}( 2\rho )}}.
\end{equation}
The above equation can be integrated analytically and the result is a linear combination of $F(\Theta|m)$ and $\Pi(n;\Theta|m)$, which are the incomplete elliptic integrals of the first kind and the third kind:
\begin{equation}
    T(\r) = \a F(\Th(\r)|m) + \b \Pi(n;\Th(\r)|m) + \tilde{t}.
\end{equation}
Here $\Th(\r)$ depends on $\r$, while $\a$, $\b$, $n$ and $m$ are constant and we set $\tilde{t} = 0$.

The derivative $\phi\pr$ of the brane scalar is given by
\begin{equation}
    \phi\prsq = \frac{1}{2}\left(H\pr +\frac{\sinh 2H}{\sinh 2\rho }\right).
\end{equation}
Using  \eqref{eq:H_sol_const_V}, we obtain the derivative of the brane scalar with respect to $t$ as
\begin{equation}
    \dot{\phi}^2 = \frac{\phi\prsq}{S^2} = \frac{2\cosh^{4} \rho }{( V\cosh( 2\rho ) -V_{0})^{2}}( V-V_{0}).
\end{equation}
Therefore, the sign of $\dot{\phi}^2$ is determined by $V-V_0$.

To determine $V_0$, we impose two kinds of boundary conditions. The first corresponds to the case where, at $t=0$, the gradient $d\r/dt=0$, i.e., at $\r=R_*$, $S(\r)$ diverges. The second is the case where, at $t=0$, the brane shrinks, i.e, $T(\r=0)=0$. The former corresponds to $R_*<R_1$, while the latter corresponds to $R_*>R_1$ in Fig.\ref{fig:phaselist}.

For the first condition, since $S(\r=R_*)$ diverges, we find
\begin{equation}
    V_0 =  V\cosh(2R_*) - \sinh(2R_*).
\end{equation}
In this case, coefficients and $\Th(\r)$ are
\begin{align}
    \a &= \frac{\sinh R_*(\cosh R_* -V\sinh R_*)}{\sqrt{V\sinh (2R_* )-\cosh (2R_* )}},&
    \b &= \frac{\sinh^{2} R_*( V-\tanh R_*)}{\sqrt{V\sinh (2R_* )-\cosh (2R_* )}},\\
    n &= \frac{1}{\cosh^2R_*},&
    m&=-\frac{(\cosh R_* -V\sinh R_*)^{2}}{V\sinh (2R_*)-\cosh (2R_*)},
\end{align}
\begin{equation}
    \cos \Theta(\r) =\frac{\sinh R_*}{\sinh \rho }.
\end{equation}
Furthermore, since $V - V_0 = 2 \sinh R_*(\cosh R_*-V \sinh R_*)$, we find that the scalar $\phi$ is real if $\cosh R_*-V \sinh R_*>0$. This result is consistent with \eqref{eq:constdphsq}.

Move on to the second boundary condition. To obtain finite $V_0$, we assume
\begin{equation}
    \frac{\sinh(2\r)}{\tanh H(\r)}\to \k < \infty.
\end{equation}
as $\r\to 0$. In other words, we assume $T(\r)$ behaves
\begin{equation}
    T(\r) \simeq \r + \frac{1-\k ^{2}}{6\k ^{2}} \rho ^{3} + O(\r^4).
\end{equation}
Note that the coefficient of $\r^3$ is always bigger than $-1/6$. Now $V - V_0 = \k$ and the corresponding coefficients and $\Th(\r)$ are
\begin{align}
    \a&=\frac{-\k V^{2} +\k +2\gamma V-2V}{2\left( \gamma -V^{2}\right)\sqrt{\k V-\gamma }},&
    \b &= \frac{(\gamma -1)(\k -2V)}{2\left( \gamma -V^{2}\right)\sqrt{\k V-\gamma }},\\
    n &= \frac{V^{2} -\gamma }{V^{2} -1},&
    m &= 2+\frac{\k V-1}{\gamma -\k V},
\end{align}
\begin{equation}
    \tan\Th(\r) = \frac{2\sqrt{\k V-\gamma }}{\k }\sinh \rho ,
\end{equation}
where we use
\begin{equation}
    \g :=\frac{\k V+1+\sqrt{\k ^{2} -2\k V+1}}{2}.
\end{equation}

\section{Constant Potential Euclidean Solutions}\label{sec:EuclidS}

Here we would like to analyze the solutions of type I and II branes in the Euclidean AdS$_3$ assuming the potential is a constant $V(\phi)=V$. To study this problem, we perform the Wick rotation of the Lorentzian global AdS$_3$ (\ref{gadsm}), by introducing the Euclidean time $\tau_E=i\t$: 
\ba
ds^2=d\rho^2+\cosh^2\rho d\tau^2_E +\sinh^2\rho d\phi^2.
\ea
The brane equation of motion can also be found by setting  $\tau=it$ in the Lorentzian one (\ref{gKEOMP}).

\begin{figure}[htbp]
    \centering
    \includegraphics[width=0.4\linewidth]{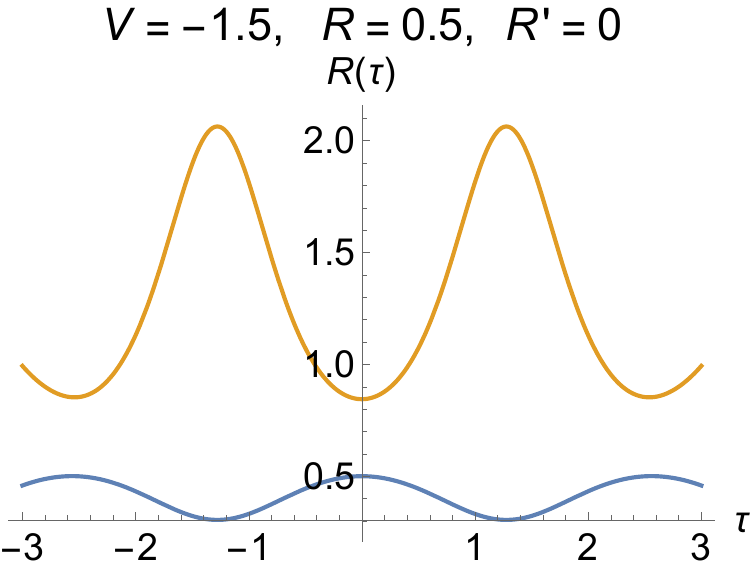}
       \includegraphics[width=0.4\linewidth]{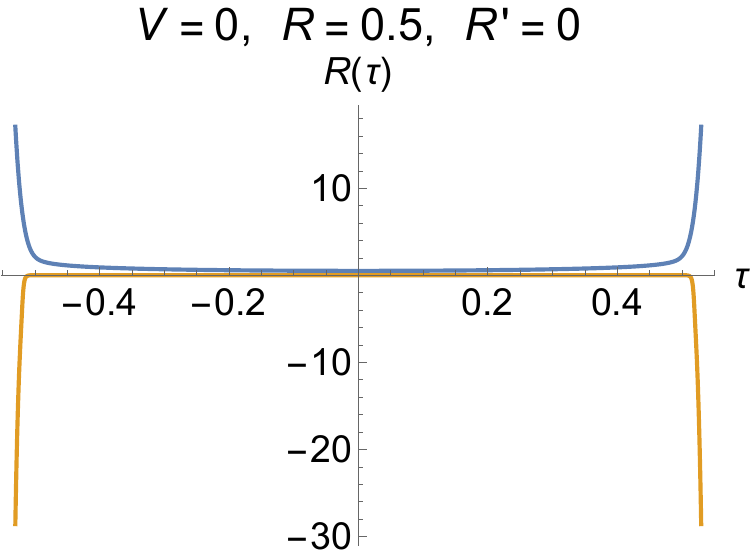}
    \caption{The plot of $R(\tau_E)$ (blue) and $\left(\frac{d\phi}{d\tau_E}\right)^2$ (orange) as a function of $\tau_E$ for the type I brane with the initial condition $R=\frac{1}{2}$ and $\dot{R}=0$ at $\t_E=0$ for $V=-1.5$ (left) and $V=0$ (right).}
    \label{fig:Ebranef}
\end{figure}

One motivation to consider the Euclidean brane solution is to set the initial condition of the Lorentzian brane solution studied in section \ref{sec:LorentzS},as we normally do in the Hartle-Hawking prescription of cosmological models. In the absence of scalar field, we have the simplest example, namely the dS$_2$ brane solution given by 
\ba
\frac{\sinh^2\rho}{\sinh ^2\xi}-\frac{\tan^2 \t}{\tanh^2\xi}=1,
\ea
where $\xi$ is a positive constant. 
We can regard this solution for $\t>0$ emerges from the semi-sphere Euclidean brane solution
\ba
\frac{\sinh^2\rho}{\sinh ^2\xi}+\frac{\tanh^2 \tau_E}{\tanh^2\xi}=1, \label{sems}
\ea
for the period $-\eta\leq \tau_E\leq 0$.

To start, let us ask if we can construct solutions with $R=0$ at a time as such a Euclidean solution provides an instanton which creates two dimensional universe.
However, a straightforward analysis of  (\ref{gKEOMP}) shows that the only possible solution looks like 
\ba
R(t)\simeq \s{\frac{2}{-\ep V}}\s{\tau_E}+O(\tau_E^{3/2}),
\ea
where $\ep=1$ and $\ep=-1$ correspond to type I and II, respectively. In this solution, the scalar field turns out to be vanishing and thus this is identical to the semi-sphere solution (\ref{sems}). Therefore we cannot construct any Euclidean instantons which create universe with a non-trivial scalar field as long as $V(\phi)$ is a constant.

Now we turn to time symmetric solutions which have a point with $\dot{R}(\equiv \frac{dR}{d\tau_E})=0$ at a specific time, which we set $\t=0$. As in section \ref{sec:LorentzS}, we define 
\ba
V=-\ep \frac{\cosh R_1}{\sinh R_1},
\ea
and $R_2=\frac{R_1}{2}$. 

\begin{figure}[htbp]
    \centering
    \includegraphics[width=0.5\linewidth]{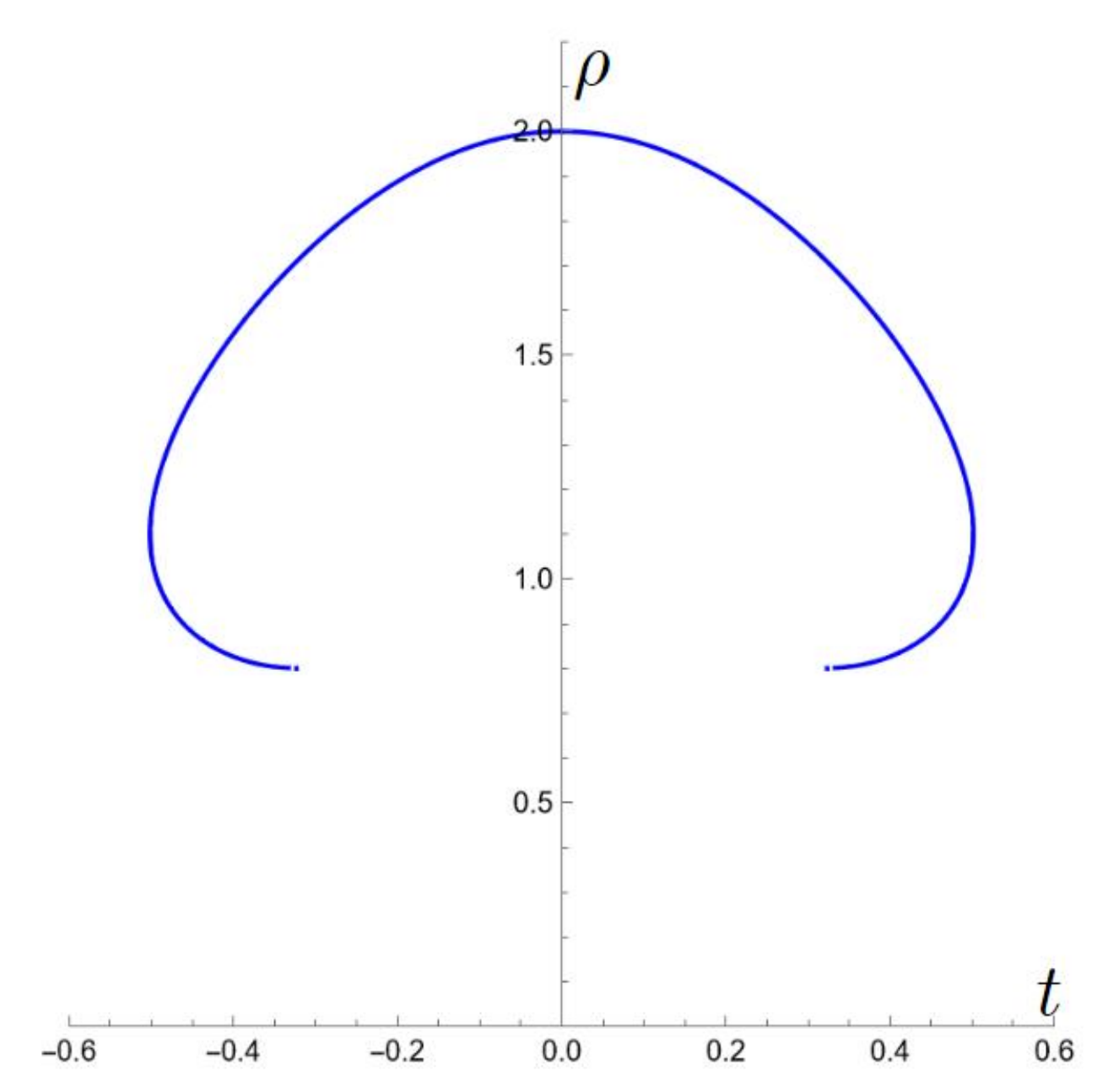}
    \caption{The profile of Euclidean type I brane at $V=-1.2$ and $R=2$. In the regime $R_*>R_1$, the EOW brane curves inwards strongly and is terminated at that point.}
    \label{fig:EulideanAnalyicalSol}
\end{figure}

First consider type I branes. When $V<-1$, we can construct oscillating solutions under $R_*<R_1$, as depicted in the left panel of Fig.\ref{fig:Ebranef}, which have the sign $\dot{\phi}^2>0$ and for $R_*>R_1$, we can only find singular solutions with $\dot{\phi}^2<0$. Fig.\ref{fig:EulideanAnalyicalSol} shows the profile of this singular brane, which is terminated in the middle. 
It is useful to note that when $\dot{R}=0$ we have $\ddot{R}>0$ (or $\ddot {R}<0$) for $R<R_2$ (or $R>R_2$) and indeed the oscillation is around $R=R_2$.
On the other hand, when $V>-1$ we can find solutions $\dot{\phi}^2>0$ with which reach the AdS boundary in the past and future as depicted in the right panel of Fig.\ref{fig:Ebranef}. Note that in order to continue to a Lorentzian solution with a real valued scalar field at $t=0$ we need $\dot{\phi}^2\leq 0 $ in the Euclidean solution. 

\begin{figure}[htbp]
    \centering
    \includegraphics[width=0.9\linewidth]{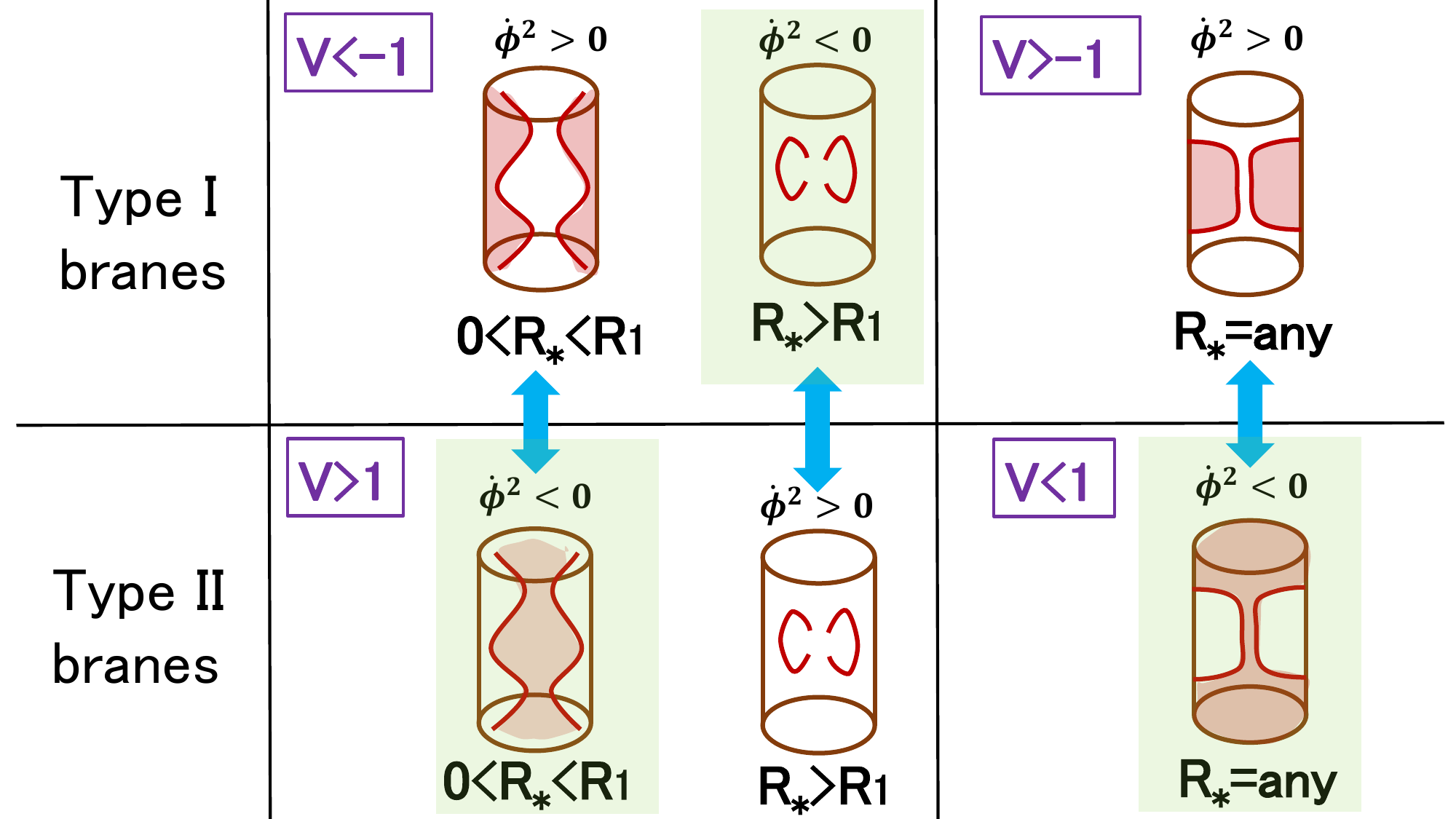}
    \caption{A sketch of profiles of Euclidean solutions in type I (upper panels) and type II branes (lower panels). The red colored regions describe the physical spacetimes inside the EOW branes. The value $R_*$ is the radius where $\dot{R}=0$.  The Euclidean solution with $\dot{\phi}^2<0$ can have a real valued scalar solutions in the Lorentzian continuation and they are colored in green. In the case $R_*>R_1$ for $V<-1$ (type I) and $V>1$ (type II), we find only singular branes which are terminated in the middle as in Fig.\ref{fig:EulideanAnalyicalSol}.}
    \label{fig:Ephaselist}
\end{figure}

To obtain results for type II branes, we can simply replace $V\to -V$ and $\dot{\phi}^2\to -\dot{\phi}^2$ and thus we omit the details. The final summary of classification of all possible solutions in type I and II branes is provided in Fig.\ref{fig:Ephaselist}.

\section{Hyperbolic Slice}\label{ap:hyp}
Consider the Hyperbolic slice in a Global AdS$_3$:
\be
ds^2=d\eta^2+\sinh^2{\eta}(-dt^2+\cosh^2{t}d\phi^2).  \label{Hypmetric}
\ee
This coordinate $(t,\eta)$ is related to the global coordinates $(\tau,\rho)$ via
\ba
\sinh{\rho}=\cosh{t}\sinh{|\eta|}, \tan{\tau}=\tanh{\eta}\sinh{t}
\ea
We insert EOW brane along the trajectory $\eta=H(t)$, which also extend in the $\phi$ direction.
Two types of global AdS$_3$ are
\ba
&& \mbox{Type I:}\ \ \ 0\leq |\eta| \leq H(t),\no
&& \mbox{Type II:}\ \ \ |\eta| \geq H(t).
\ea
We can write induced metric as follows:
\ba
ds^2=-(\sinh^2{H}-\dot{H}^2)dt^2+\sinh^2{H}\cosh^2{t} d\phi^2,
\ea
where $\dot{H}=\frac{dH(t)}{dt}$. Since we assume that the EOW brane is time-like in the bulk AdS$_3$, we require $\sinh^2 H>\dot{H}^2$. Since induced metric is diagonal, the null energy condition is also $\dot{\phi}^2\geq0$.

We choose the normal vector to be
\ba
N_a=\eps \frac{\sinh H}{\s{\sinh^2 H-\dot{H}^2}}\left(1,-\dot{H},0\right),
\ea
where $\epsilon=\pm 1$. When $\eps=1$ (or $\eps=-1$), $N$ has a positive (or negative) direction on the $z$ axis, which is a type I (or type II) set up.

This leads to 
\ba
&K_{tt}=\eps \frac{2\dot{H}^2\cosh{H}-\sinh{H}(\cosh{H}\sinh{H}+\ddot{H})}{\s{\sinh^2 H-\dot{H}^2}}, \\
&K_{\phi\phi}=-\eps \frac{\cosh{t}\sinh{H}(\cosh{t}\cosh{H}\sinh{H}+\dot{H}\sinh{t})}{\s{\sinh^2 H-\dot{H}^2}},\\ &K_{t\phi}=0.
\ea
The boundary condition (\ref{KEOM}) can be written as follows:
\ba
&\dot{\phi}^2=\eps \frac{\sinh H \left(-\ddot{H}+\dot{H}\tanh t +\dot{H}^2 \coth {H}-\dot{H}^3\tanh t 
   \text{csch}^2{H} \right)}{2 \s{\sinh ^2{H}-\dot{H}^2}},\label{HypKEOMP} \\
&V(\phi)= \eps \frac{\left( \sinh{H} \left(\ddot{H}-\dot{H} \left(3\coth{H}+\dot{H}\text{csch}^2H \tanh{t} \right)+ \sinh 2H \right) \right)}{2 \left(\sinh^2 H-\dot{H}^2 \right)^{3/2}}.
\label{HypKEOMQ}
\ea

We obtain a simple example, $V(\phi)=\mathrm{const}$. We choose
\be
H(t) = \xi.
\ee
The boundary conditions (\ref{HypKEOMP}, \ref{HypKEOMQ}) are written as follows;
\ba
&& \dot{\phi}^2 = 0 \\
&& V(\phi) = \coth{\xi} \label{HypsimpleV}
\ea
Thus the behavior of the scalar field is
\be
\phi=\mathrm{const.}
\ee
This is the important example because it can be both  type I and type II.
%%%%%%%%%%%%%%%%%%%%%%%%%%%%%%%%%%%%%%%%%%%%%%%%%%%%%%%%
%%%%%%%%%%%%%%%%%%%%%%%%%%%%%%%%%%%%%%%%%%%%%%%%%%%%%%%%
% bibliography via BibTeX
\bibliographystyle{JHEP}
\bibliography{dSBCFT}

%%%%%%%%%%%%%%%%%%%%%%%%%%%%%%%%%%%%%%%%%%%%%%%%%%%%%%%%
%%%%%%%%%%%%%%%%%%%%%%%%%%%%%%%%%%%%%%%%%%%%%%%%%%%%%%%%

\end{document}